*Review*

# Possibilities for an Aerial Biosphere in Temperate Sub Neptune-Sized Exoplanet Atmospheres

Sara Seager [1,2,3,*], Janusz J. Petkowski [1], Maximilian N. Günther [2,†], William Bains [1,4], Thomas Mikal-Evans [2] and Drake Deming [5]

1. Department of Earth, Atmospheric, and Planetary Science, Massachusetts Institute of Technology, Cambridge, MA 02139, USA; jjpetkow@mit.edu (J.J.P.); bains@mit.edu (W.B.)
2. Department of Physics, and Kavli Institute for Astrophysics and Space Research, Massachusetts Institute of Technology, Cambridge, MA 02139, USA; maxgue@mit.edu (M.N.G.); tmevans@mit.edu (T.M.-E.)
3. Department of Aeronautics and Astronautics, Massachusetts Institute of Technology, Cambridge, MA 02139, USA
4. School of Physics & Astronomy, Cardiff University, 4 The Parade, Cardiff CF24 3AA, UK
5. Department of Astronomy, University of Maryland at College Park, College Park, MD 20742, USA; lddeming@gmail.com
* Correspondence: seager@mit.edu
† Juan Carlos Torres Fellow.

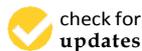

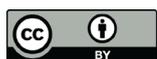

**Abstract:** The search for signs of life through the detection of exoplanet atmosphere biosignature gases is gaining momentum. Yet, only a handful of rocky exoplanet atmospheres are suitable for observation with planned next-generation telescopes. To broaden prospects, we describe the possibilities for an aerial, liquid water cloud-based biosphere in the atmospheres of sub Neptune-sized temperate exoplanets, those receiving Earth-like irradiation from their host stars. One such planet is known (K2-18b) and other candidates are being followed up. Sub Neptunes are common and easier to study observationally than rocky exoplanets because of their larger sizes, lower densities, and extended atmospheres or envelopes. Yet, sub Neptunes lack any solid surface as we know it, so it is worthwhile considering whether their atmospheres can support an aerial biosphere. We review, synthesize, and build upon existing research. Passive microbial-like life particles must persist aloft in a region with liquid water clouds for long enough to metabolize, reproduce, and spread before downward transport to lower altitudes that may be too hot for life of any kind to survive. Dynamical studies are needed to flesh out quantitative details of life particle residence times. A sub Neptune would need to be a part of a planetary system with an unstable asteroid belt in order for meteoritic material to provide nutrients, though life would also need to efficiently reuse and recycle metals. The origin of life may be the most severe limiting challenge. Regardless of the uncertainties, we can keep an open mind to the search for biosignature gases as a part of general observational studies of sub Neptune exoplanets.

**Keywords:** exoplanets; exoplanet atmospheres; biosignature gases

## 1. Introduction

We aim to expand our understanding of what kind of exoplanets have habitable environments. We begin with the motivation to consider the atmospheres of a type of exoplanet called "sub Neptunes" (Section 1.1). We then summarize Earth's atmospheric aerial biosphere (Section 1.2), followed by a review of theoretical studies on giant planet and brown dwarf atmospheres as potential abodes for life (Section 1.3).

*1.1. Motivation to Expand the Pool of Planets Considered to Be Potentially Habitable*

The growing excitement of the chance to discover signs of life on exoplanets by way of exoplanet atmosphere biosignature gases is motivating many new simulations to assess detectability. The excitement is tempered by the reality that only a small number of potentially habitable rocky planets will be accessible for such observations with the next-generation telescopes expected in the next two decades, including the James Webb Space Telescope (JWST) [1], the "extremely large" ground-based telescopes now under construction [2–4], and possible space observatories such as the Starshade Rendezvous Probe [5], The HabEx Observatory [6], the LUVOIR telescope [7], or the Origins Space Telescope [8].

The "traditional" habitable planet is rocky, with a thin atmosphere, most usually considered to have a high mean molecular weight atmosphere, such as one dominated by $N_2$ or $CO_2$. Such habitable planets, however, present formidable challenges in detecting the components of their thin atmospheres. If planets more favorable for observational study than small, rocky exoplanets are valid targets for the detection of signs of life, then we need to expand our understanding of the habitable planet parameter space.

In this paper, we focus on the temperate sub Neptune-sized exoplanets (hereafter called temperate sub Neptunes) as candidate abodes for life. Sub Neptunes are loosely defined as planets larger than Earth but smaller than Neptune, with radii between 1.6 and 3 or 4 Earth radii ($R_{Earth}$). Sub Neptunes are distinguished from 'super-Earths', which are also planets larger than Earth but smaller than sub Neptunes, by having thick envelopes of $H_2$ or $H_2$-He that comprise 1 to 10% of the planetary mass [9,10]. We further define "temperate" to highlight the sub Neptunes that receive Earth-like amounts of energy from their host star (e.g., K2-18b [11]), because of such planets' special potential to have liquid water either in cloud form [12] or even as an ocean beneath the atmosphere [13,14].

Sub Neptunes are favorable for both detection and atmospheric study because of their large sizes and lower densities as compared to rocky exoplanets [15]. Those transiting red dwarf stars have larger transit signals as compared to those transiting sun-sized stars. Hundreds of sub Neptunes are known (Figure 1), such that they appear to be the most common type of planet in our Galaxy (out to orbital periods of 100 days) [16–18].

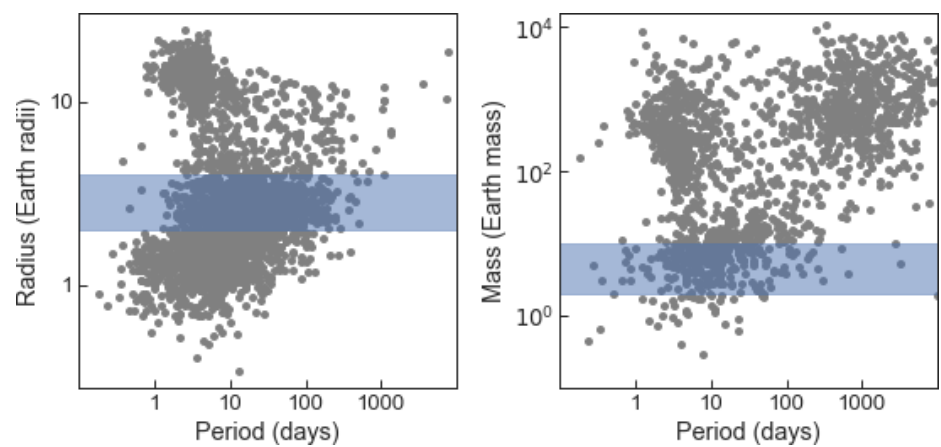

**Figure 1.** The known population of exoplanets with the sub Neptune exoplanets marked by the blue bands. The left (right) panel shows the planet radius (mass) in Earth radii (mass) versus orbital period in days. The blue-shaded region marks the typically adopted sub Neptune defining range of approximately 1.6–4 Earth radii (2 to 10 Earth masses). Only confirmed exoplanets are plotted, hence the data shown reflect selection and detection biases. Each panel shows slightly different samples of exoplanets because some planets have only mass or radius measured and not both. Data from Exoplanet archive (https://exoplanetarchive.ipac.caltech.edu/ (accessed on 28 January 2021)).

Despite their common occurrence, the sub Neptunes' bulk interior composition remains a mystery. Nonetheless, there are a number of possibilities ([13,19], Figure 2). Most sub Neptunes need an extended gas envelope to explain their low densities [10]. Sub Neptunes may be rocky cores surrounded by an outgassed hydrogen ($H_2$) envelope, or a hydrogen and helium (He) envelope captured from the protoplanetary nebula. Such planets may have low water content atmospheres, or alternatively, a substantial interior water layer and high water atmosphere content, if, for example, the planet formed beyond the snow line where water ice is a common planetary building block. A more extreme water content version is the concept that sub Neptunes are water worlds, composed of 50–90% water by mass. Water world sub Neptunes may even harbor a liquid water ocean if the interior temperature is low enough [13,14]. For a recent review of sub Neptunes, their formation, and their possible interiors, see [19].

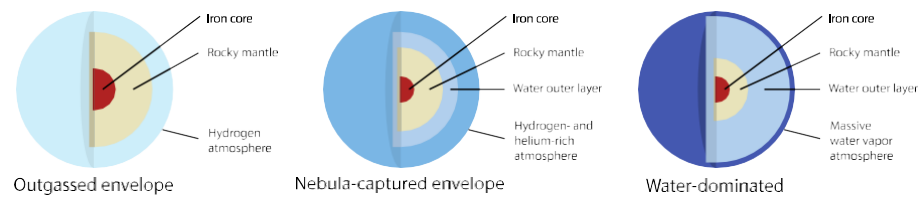

**Figure 2.** Schematic of concepts for sub Neptune interior and envelope composition. *Left*: A sub Neptune may have a massive core and mantle surrounded by an outgassed $H_2$ envelope with low atmospheric water vapor content. *Middle*: A sub Neptune may have a core and mantle surrounded by a water layer itself covered by an $H_2$–He envelope. *Right*: A water world composed of 50–90% water by mass will have a water-dominated atmosphere and interior. The inner water layer may include a water ocean, a supercritical water layer, and high-pressure water ice. Adapted from [13].

The key point is that life for the sub Neptunes considered here would have to exist in the atmosphere. Sub Neptunes have no solid or liquid surface similar to Earth's surface [20]. Far beneath the sub Neptune atmosphere, ($10^3$ to $10^4$ km below [21]), there may be a rocky core with a solid outer layer with significant volcanic activity, or even a magma ocean [21,22]. However, the temperature at any core–envelope boundary (~2000–3000 K [21–23]) is too hot for any complex molecules to be stable, and hence for life of any kind. Furthermore, the high temperatures and high pressures (1 to 10 GPa [21]) mean any surface does not resemble the terrestrial planet or moon surfaces we are familiar with and will be materially quite different. The lack of a temperate, rocky surface means that life on a sub Neptune would have to originate in the clouds (by chemistry or interplanetary transmission) and perpetually survive in the atmosphere.

By life in the atmosphere, we mean microscopic life in the form of small life particles. This life would be microbial-like, although we do not imply that hypothetical "bacteria" might in any way be taxonomically related to microorganisms on Earth. The point is life has to be small enough to maintain a decent residence time in the atmosphere without sinking out due to gravity. Such life could be simple, single-celled life and passively move about the atmosphere. Or the life form might be more complex, perhaps even capable of active flight. For life to generate a biosignature gas, the life must use chemistry as life on Earth does, to extract, store, and use energy for metabolism, and to generate a waste gas that is a distinctive spectrally and present in high enough quantities to conceivably be detected. Life on Earth generates gaseous waste products and it is reasonable to assume that life elsewhere would do so as well [24]. Our assumptions do not require life elsewhere to have the same chemistry as life on Earth, although we tacitly assume throughout that life is based on carbon and water, as there is widespread consensus that this is most likely [25].

For practical purposes, we focus on planets that are observable in this decade, which limits us to temperate sub Neptunes transiting M dwarf stars. Such planets can be discovered with, e.g., TESS [26], and their atmospheres observed with Hubble or JWST. Because transit probability scales as $R_*/a$, where $R_*$ is the stellar radius and $a$ is the planet orbital semi-major axis, planets transiting sun-like stars or cold sub Neptunes far from M dwarf stars are too rare to be likely observational targets. However, our work can be extended to temperate sub Neptunes orbiting sun-like stars and cold planets orbiting both M dwarf stars and sun-like stars in the future when ground- and space-based high-contrast direct imaging, a technique not limited to the rare transiting planets, becomes a reality [5–7].

### 1.2. Earth's Aerial Biosphere

Before postulating an aerial biosphere on temperate sub Neptune exoplanets, it is worthwhile to review life in Earth's atmosphere, for context. Earth has an aerial biosphere created by microbial life that regularly migrates to clouds from the ground [27,28]. Earth's cloud aerial biosphere is thought to serve as a temporary refuge during long distance transportation across whole continents and oceans [29–31], with microbes remaining aloft on average for 3 to 7 days [32]. Microbes are eventually deposited to the surface by precipitation [27].

Earth's aerial biosphere has microbial concentrations ranging from $10^3$ m$^{-3}$ to more than $10^6$ m$^{-3}$ in air. Approximately 20% of the microbes are larger than 0.5 μm in diameter [33]. For reference purposes, we can show that on Earth, the average concentration of cells in collected cloud liquid water is comparable to that in open ocean waters. Earth's liquid water clouds have $1 \times 10^5$ cells mL$^{-1}$, as measured in collected cloud liquid water [34] (i.e., $3 \times 10^4$ cells m$^{-3}$ of air that includes cloud liquid droplets), whereas the open ocean has $5 \times 10^5$ cells mL$^{-1}$ in the upper 200 m column, and $5 \times 10^4$ cells mL$^{-1}$ below 200 m depth [35]. For a detailed summary of Earth's aerial biosphere, see [36] and references therein.

Earth's clouds are not a permanent habitat for life (e.g., [37]). The clouds are a challenging ecological niche for permanent habitation because of their transient and fragmented nature. The additional fact that Earth's surface is habitable means that there is limited evolutionary pressure to overcome substantial barriers for terrestrial life to evolve a lifecycle permanently sustained in the clouds.

The microbes mostly reside inside cloud droplets but some are free-floating in the atmosphere. Some microbes (both inside and outside droplets) are found to be metabolically active. Although there is not yet direct evidence of cell division in situ in the clouds [38], the possibility for active cell division remains viable [39–41].

The microbial life in Earth's atmosphere is diverse, including bacteria, archaea, eukaryotes, and viruses [28,38], and shows a surprisingly varied set of physiologically active metabolisms in cloud droplets [38]. Diverse physiological and biochemical strategies of microbes have been identified that seem to be direct, specific adaptations to the cloud droplet environment. The adaptations include protection against oxidants, osmotic pressure variations, the synthesis of cryoprotectants to fight extreme cold, or production of metal ion scavengers and biosurfactants [28,38].

### 1.3. Speculation on Solar System Planet and Brown Dwarf Aerial Biospheres

Aerial biospheres may not be limited to Earth. Aerial biosphere hypotheses for other planetary bodies include the Venusian atmosphere—a topic speculated on for many decades. (For the original idea, see [42] and for a recent review, see [43].) Venus' atmosphere has a cloud-filled layer that is at a suitable temperature for life. The clouds permanently cover the entire planet and the temperate layers span a large vertical extent from 48 to 60 km altitude. The Venus cloud environment is undeniably harsh (for a review, see [36]), both very dry and very acidic. The clouds are composed of liquid sulfuric acid, with acidity more than ten orders of magnitude more acidic than the most acidic environment on Earth [44,45].

If populated by life, how the Venusian clouds could have become inhabited is unknown (see [36]). Perhaps the most conceivable scenario is that life originated on the surface when Venus was colder and surface temperatures were cool enough to support a water ocean—and that life only permanently migrated into the clouds after Venus' greenhouse runaway caused surface temperatures to become inhospitable. Today, Venus' surface at 735 K is too hot for life of any kind, because the surface temperatures are too high for any plausible solvent and for most organic covalent chemistry.

Nonetheless, [36] propose a Venus life cycle whereby: (1) spores populate the Venus atmosphere lower haze layer, a stagnant layer that is little understood; (2) desiccated spores travel up by mixing via gravity waves, followed by convective entrainment; (3) spores act as cloud condensation nuclei and, once surrounded by liquid germinate and become metabolically active; (4) metabolically active microbes grow and divide within liquid droplets, the liquid droplets grow by coagulation; (5) after months or years, the droplets become large enough to gravitationally settle down out of the atmosphere; droplet evaporation triggers cell division and sporulation; and back to (1). The spores are small enough to withstand further downward sedimentation, remaining suspended in the lower haze layer "depot".

Jupiter's atmospheric cloud layers have also been considered as a habitat for life, as the Jovian atmosphere has a temperate cloud layer. Any life form, however, would be subject to downward motion by gravitational settling, convective downdrafts, or meridional overturning to layers with destructively high temperatures. There is no stagnant layer in Jupiter's atmosphere as there is on Venus, so the chance for life to exist by avoiding the hottest atmosphere layers appears less favorable on Jupiter than on Venus. In general, on Jupiter, and any giant planets, beneath the extended $H_2$/He envelope, the planet is far too hot for life (as there is no temperate surface as a barrier for life's descent).

Nonetheless, [46] argue for four different ecological niches in the Jovian atmosphere. This includes postulating that life forms that might grow fast enough to replicate before being drawn down to hot layers, or that life may have an active mechanism for staying aloft (including slow powered locomotion or buoyancy control). One of Sagan's speculative life forms, "balloon organisms" [46], is biologically unrealistic in Jupiter's atmosphere that is predominantly $H_2$ and He, mainly because getting enough lift would require a very large internal reservoir of almost entirely pure $H_2$. Keeping a pure $H_2$ internal gas reservoir, or alternatively heating the internal gas, is prohibitively energetically costly for a life form.

More recently, [47] considered life in the coolest free-floating brown dwarfs ("Y dwarfs"), objects with cool enough atmosphere layers for liquid water clouds. The motivation to study Y dwarfs is that there are likely tens of Y dwarfs within 10 pc (33 light years) of Earth and billions more throughout our Galaxy [47]—possibly making the Y dwarfs the most abundant sites for life. However, even the coolest brown dwarfs have the same fundamental limitation to life that Jupiter's atmosphere does: beneath any temperate atmosphere layers is a vast, destructively hot planetary atmosphere and interior. Inspired by [46,47] modeled organisms as individual frictionless hollow spheres with a permeable skin. They focus on a 1D convection and gravitational settling lifecycle model to constrain life particle sizes, by calculating organism sizes and masses that can float indefinitely in a convective updraft of a given speed. They find that microbes could indeed persist in the Y dwarf atmosphere convective bands (i.e., latitudes), and for atmospheres with convection, microbes could be up to an order of magnitude larger than typical microbes on Earth. Ref. [48] further the case for brown dwarf habitability by describing the habitable volume in brown dwarf atmospheres, as well as describing prebiotic chemistry, abiogenesis, nutrient and energy supplies, and observational prospects.

In this paper, we synthesize and build upon existing but fragmented building blocks to assess the plausibility of life in sub Neptune atmospheres. We begin by describing how temperate sub Neptune atmospheres may satisfy the basic requirements of life (Section 2). We next show how only those atmospheres with high water content or cool lower layers can support liquid water clouds (Section 3). We qualitatively describe how life might persist aloft amidst fragmented clouds with no temperate surface below (Section 4). We then summarize additional challenges for life in an aerial biosphere (Section 5). We speculate on possible metabolisms and resulting biosignature gases (Section 6). We round out our review with a list of temperate sub Neptune exoplanets suitable for atmosphere observations (Section 7). We conclude with a summary (Section 8).

## 2. Sub Neptune Atmospheres Can Satisfy the Basic Requirements for Life

Theory, data, and experiments suggest that the requirements for life, in decreasing order of certainty, are [25]:

- A thermodynamic disequilibrium;

- An environment capable of maintaining covalent bonds (in chemical compounds), especially between carbon, hydrogen, and other atoms;
- A liquid environment;
- A molecular system that can support Darwinian evolution.

Here, we qualitatively describe how sub Neptune atmospheres can satisfy each basic requirement for life.

*Thermodynamic disequilibrium.* Sub Neptune atmospheres have stellar radiation as a source of thermodynamic disequilibria. Sub Neptune atmospheres may also have chemical potential energy gradients, driven in the atmosphere by photochemistry or by vertical transport moving chemical species from the deep atmosphere or interior to the upper atmosphere. It is worth noting, however, that many sources of thermodynamic disequilibria present on rocky planets are absent, including, for example hydrothermal vents.

*An environment capable of maintaining covalent bonds.* Temperate sub Neptune atmospheres can have the right temperatures and pressures to maintain covalent bonds, needed for large, diverse molecules. For example, different models for K2-18b estimate atmospheric temperatures at pressures above 0.1 bar, ranging from 250 to 400 K [12,14,49–51]. Note that pressure will have only a minor role on the chemical stability of metabolism compared to the effects of temperature [52].

Regarding temperature limits for Earth life, the lower temperature limit is 258 K (microbes with cold tolerance) and the upper limit for life has been suggested as between 395 K [53] (based on protein degradation rates) and 425 K (based on metabolite hydrolysis rates) [52]. The most heat-tolerant organisms found can grow at 395 K [54,55].

*A liquid environment.* Some temperate sub Neptunes may have liquid water clouds in their atmospheres. With elemental hydrogen and oxygen, water vapor will be present in sub Neptune atmospheres. The atmosphere, however, needs the right temperature and pressure for water to condense to liquid instead of ice (Section 3).

*A molecular system that can support Darwinian evolution.* There is nothing in the sub Neptune atmosphere that precludes existence of a molecular system that can support Darwinian evolution. The ability for life to have chemical complexity is dependent on three general criteria: sufficient chemical diversity, chemical reactivity, and the presence of a solvent (reviewed in [56,57]).

In addition to the above four criteria, we must also emphasize that the bulk gases in sub Neptune atmospheres—$H_2$ and He gases or $H_2O$ vapor for water worlds—are not detrimental to life. $H_2$ and He are relatively inert when it comes to life processes. Indeed, life can survive, reproduce, and thrive in an $H_2$- or $H_2$ and He-dominated environment, with no adverse effects [58].

## 3. Atmospheric Liquid Water Clouds Require High Water Content or Cold Lower Layers

Life requires a liquid environment in order to provide a solvent, i.e., a medium in which basic biochemical reactions can occur. For the possibilities of an aerial biosphere, we are interested in the conditions for which liquid water clouds can form in the atmosphere. For this paper we are focusing on temperate sub Neptunes orbiting M dwarf stars the most favorable case for current detection and study.

We can use the water equilibrium phase diagram and assume that condensation and cloud formation may take place whenever the partial pressure of the vapor exceeds the saturation vapor pressure given by the condensation curves shown in Figures 3 and 4. We stress that the temperatures and pressures for which water can be in the liquid phase changes depending on the total water content on an atmosphere (Figures 3 and 4, and, e.g., [59]). It is key to recall the importance of the water content of the atmosphere: at a fixed temperature a lower water content requires a higher pressure to condense [59]. We can think of this colloquially as a higher abundance of water molecules per unit volume promotes their condensation.

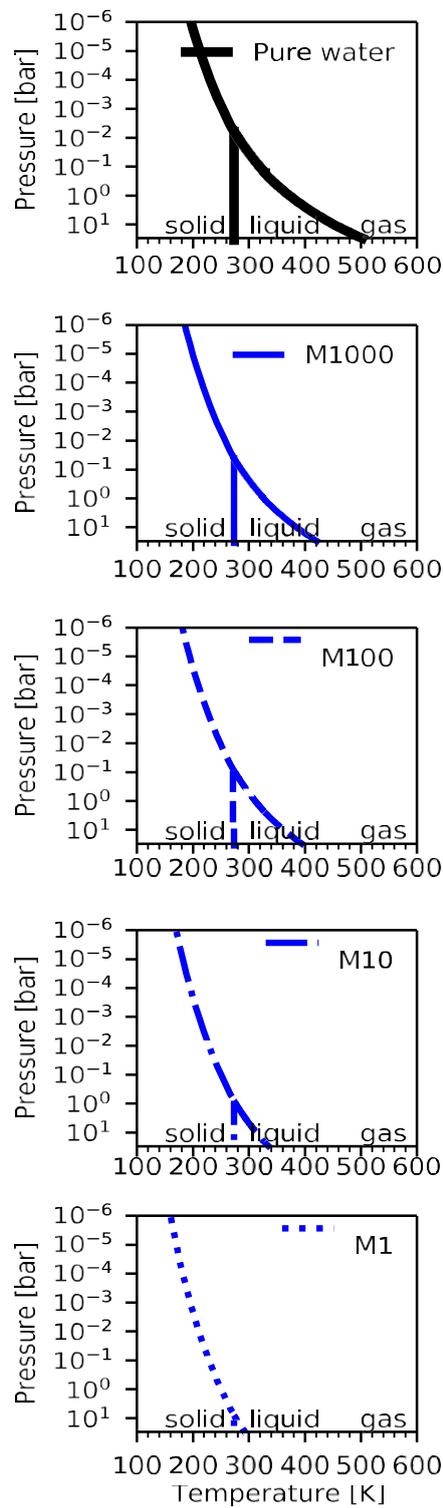

**Figure 3.** Water equilibrium phase diagram showing condensation curves for varying water content. The *x* axis shows temperature and the *y* axis shows pressure. Metallicity values translate to approximate water vapor content in a 90% $H_2$ and 10% He atmosphere as follows: M1 (solar metallicity) ~ 0.00078; M10 (10× solar) ~ 0.0077; M100 (100× solar) ~ 0.068; M1000 (1000× solar) ~ 0.16. The lines show the condensation curves. The blue-shaded region shows the temperature and pressure range for the liquid water phase. The series of plots shows that, for a fixed temperature, and compared to the pure water phase case, water condenses at lower pressures for lower water content. The pressure for which water can be in liquid phase changes for differing atmospheric total water content.

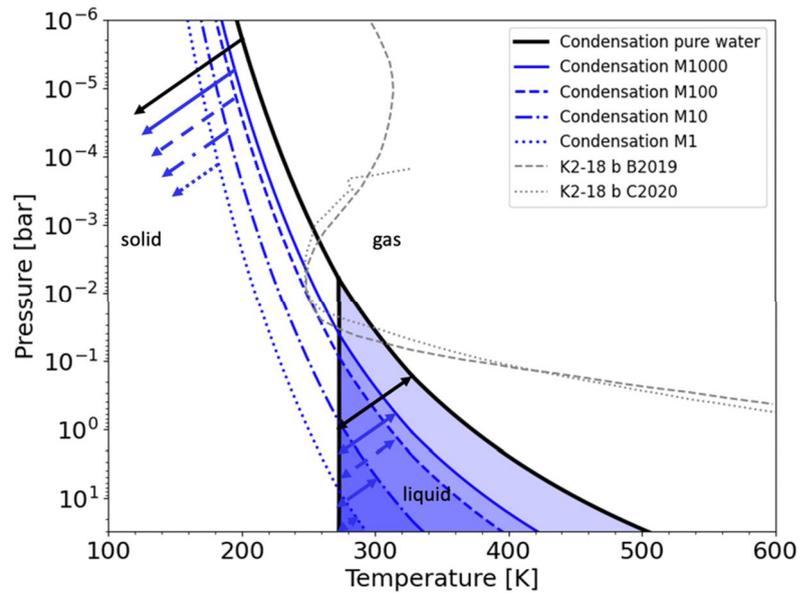

**Figure 4.** Water equilibrium phase diagram showing condensation curves for varying water content. The *x* axis shows temperature and the *y* axis shows pressure. The black solid line is the condensation for pure water and the blue lines show condensation for atmospheres of different water content (corresponding to the different panels in Figure 3). Metallicity values translate to approximate water vapor content in a 90% H$_2$ and 10% He atmosphere as follows: M1 (solar metallicity) ∼ 0.00078; M10 (10× solar) ∼ 0.0077; M100 (100× solar) ∼ 0.068; M1000 (1000× solar) ∼ 0.16. The arrows in the top left show that to the left of each condensation curve water will be in solid form (ice). The blue-shaded regions indicate the temperatures and pressures where water is in the liquid phase for different water content; the double-headed arrows indicate the liquid water range for different values of atmospheric water content. The black vertical line is the liquid/solid phase boundary common to all values of atmospheric water content, because the freezing temperature for water (273 K) is insensitive to pressure for pressures less than approximately 100 bar. The grey dashed and dotted lines show two different K2-18 b atmosphere profiles that fit Hubble data (from [12,49]), both are for atmospheric metallicities of 100× solar (B2019 is the Bond albedo = 0.3 case from [12] and C2020 is from [51]). Because the grey curves intersect the condensation curve for an atmosphere with metallicity of 100 (blue dashed curve) but do not pass through the liquid water region for an atmosphere metallicity of 100, the models show that water will be in ice and not liquid form. To support liquid water clouds, planet atmosphere temperature–pressure profiles must pass through the liquid water region of the phase diagram.

Relevant to a sub Neptune's atmospheric water content is that astronomers use the term metallicity to describe the enrichment of any element heavier than helium relative to solar composition. Metallicity is conventionally defined as $\left[\frac{Fe}{H}\right] = \log_{10}\left(\frac{N_{Fe}}{N_H}\right)_{planet} - \log_{10}\left(\frac{N_{Fe}}{N_H}\right)_{Solar}$, where $N_{Fe}$ and $N_H$ are the number density of iron and hydrogen atoms (i.e., per unit of volume), respectively, the subscript planet refers to planet atmosphere values and the subscript solar is for solar values. Jupiter, Saturn, Uranus, and Neptune are increasingly metal rich [60]. It is unknown whether their metallicity is due to planet mass or formation location in the solar nebula. Hence, we must await observations of sub Neptunes to determine the metallicity of both individual planet atmospheres and the population as a whole.

We now turn to the focus of this section, to review what water content can lead to liquid water clouds in temperate sub Neptunes orbiting M dwarf host stars. For example, for an atmospheric water vapor abundance of approximately 6 parts per ten thousand (corresponding to approximately solar metallicity for an atmosphere composed of 90% H$_2$ and 10% He), water will exist as a liquid only at pressures higher than approximately 10 bar. At 10 bar atmospheric pressure, the vapor pressure of water would be 0.006 bar, which is the vapor pressure of water over liquid at 273 K. If the pressure were lower than 10 bar or the temperature higher than 273 K, any liquid water would evaporate. If the temperature were lower than 273 K, then water would condense as ice (note that the freezing point of water (273 K) is independent of pressure for pressures less than 100 bar). We should note that 10 bar is generally too high of a pressure to be probed by observations, so for a solar metallicity atmosphere of a temperate sub Neptune, the concept of an aerial biosphere might not be a useful one. (It is also useful to note that according to Dalton's Law, for an ideal gas the molar concentration is the ratio of partial pressure to total pressure.)

The planet's atmosphere temperature–pressure profile must go through the liquid region of the water phase diagram for liquid water clouds to exist. One possibility is high water content in the atmosphere, which for a sub Neptune corresponds to high metallicity (see the liquid phase region of the water phase diagram in Figures 3 and 4). Atmosphere models for the sub Neptune K2-18b, currently the only temperate sub Neptune with atmosphere observations, show that only water ice clouds and not liquid water clouds can form, given the exterior heating from the host star and the internal heating from interior energy [50,51]. This important point is made by [50,51], who show that K2-18 b has to have high water content, with a metallicity of greater than 100x solar, in order for water to condense, and that water condenses to the ice phase and not the liquid phase (as seen in Figure 4). Other than high water content, sub Neptunes must have lower interior energies than assumed for current K2-18 b models in order for liquid water clouds to exist, according to Figure 4. For a detailed description of various sub Neptune 1D atmosphere models that do support liquid water, see [61].

## 4. How Can Life Persist Aloft?

A key question for an aerial biosphere on a sub Neptune, or any planet without a temperate, solid surface, is, "how can life persist aloft in the atmosphere indefinitely?" All planet atmospheres ultimately get warmer with decreasing altitude, so that, if there is not a temperate surface as a barrier to descent, a descending organism will reach an environment where the planet's temperatures will be too hot for life of any kind to survive. Life particles are subject to downward motion by winds, convection, atmospheric circulation, or gravitational settling for large particles. If the planet is to be inhabited, life must have enough time aloft in order to reproduce and spread, before downward transport, in order for life to maintain itself at a habitable altitude. Alternatively, the lowest levels of the atmosphere that a life particle reaches must be within a suitable temperature for life. Life must furthermore spend enough time aloft in a liquid environment, so that life can metabolize and reproduce. Yet, cloud formation and extent is a complex topic. We review how passive microbial-type life particles might be supported and suggest helpful avenues for future work.

### 4.1. Can Passive Life Particles Spend Enough Time Amongst Liquid Cloud Droplets?

Passive microbial-type life particles must spend enough time in a liquid water cloud region even if they are not residing inside water droplets (see Section 4.1.3). Ideally, we would use Lagrangian particle dynamics and atmospheric circulation models to follow the trajectory of airborne microbial-type particles through temperate sub Neptune exoplanet atmospheres. In this way, we could assess how long particles spend in liquid water cloud locations and how often many of the particles escape downward transport to destructively hot atmospheric layers. In lieu of such models, we qualitatively describe the possible journey of life particles, using results from the one existing paper on temperate sub Neptune planet atmospheric circulation, a pioneering treatment of K2-18b [51]. We next expand our discussion to planets more rapidly rotating than K2-18b.

4.1.1. Cloud Location and Downward Transport in a Slow Rotator: K2-18b

The main characteristics of K2-18b as related to atmospheric dynamics are its slow rotation rate (32 days, assuming K2-18b is tidally-locked such that its rotation rate matches its orbital period), its radius (2.6 $R_{Earth}$, and its moderate stellar heating (close to the insolation received by Earth). These factors imply weak horizontal temperature gradients and a single dayside/nightside overturning circulation (see, e.g., the Weak Temperature Gradient approximation in [62]). Indeed, the atmospheric circulation models of K2-18b [51] show a single planet-wide circulation cell with a symmetric day–night circulation and efficient heat redistribution (Figure 5a). Atmospheric mass moves upwards on the dayside surrounding the substellar region, flows away from dayside around the planet, and downwells at the nightside. The atmosphere recirculates back to the dayside at a low atmospheric altitude (Figure 5).

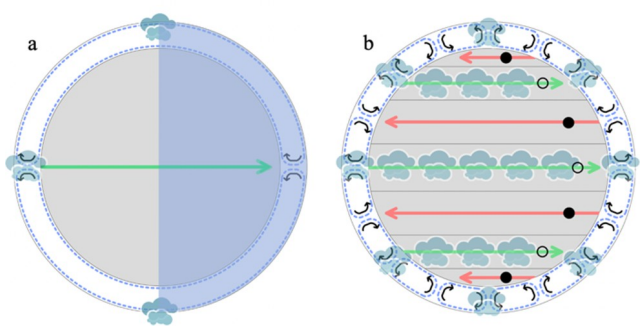

**Figure 5.** Qualitative illustration of the atmospheric circulation for and preferential location of cloud formation on slowly rotating and rapidly rotating sub Neptunes (not to scale). (**a**) A tidally-locked temperate sub Neptune will be slowly rotating and is expected to have a single planet-wide circulation cell with a symmetric day–night circulation moving away from the substellar point (left side of the figure at the base of the green arrow) and returning at a lower level after downwelling at the night side. Clouds may form at the dayside (left side of the figure) and/or terminator, depending on a complex interplay of a variety of factors and based on simulations [51]. The blue-shaded region indicates the permanent planet night side. Adapted from [51]'s description of K2-18 b. (**b**) A rapidly rotating planet is expected to have multiple latitudinal cells, with alternating wind direction (indicated by arrows). The latitudinal cells may also alternate between upwelling regions (indicated by open circles with dots) with clouds and downwelling regions (indicated by filled circles) that are cloud free, though this picture will be complicated by other dynamics. A key takeaway is both the slowly and rapidly rotating planets likely have a return atmosphere flow at a low atmosphere altitude; if the temperature is cool enough life can survive indefinitely.

Also important for temperate sub Neptunes is that overturning large-scale convection is inhibited when the condensable species (in this case $H_2O$) is heavier than the background gas ($H_2$) [63,64]. This is an advantage for microbes to be passively maintained aloft, in that there will not be vigorous downward transport by convection of life particles to destructively hot layers.

To describe what might happen to airborne microbial-type life particles, we choose the 300x solar metallicity atmosphere with a cloud condensation nuclei (CCN) population of $10^5$ kg$^{-1}$ illustrated in [51]. We consider life particles within 30 degrees of the K2-18b equator (mean equatorial conditions from Figure 5 in [51]).

The water cloud is confined to the day side substellar area but is large in extent. On average the cloud ranges from -40 degrees to 40 degrees longitude. The water cloud vertical extent is equally impressive, approximately 40 km considering the pressure range of approximately $10^{-2}$ to 7x$10^{-3}$ bar and converting to distance using the pressure scale height of approximately 30 km for the 300x solar metallicity atmosphere.

We are interested in the time life particles spend within a water cloud, most desirably inside liquid droplets but also free floating outside droplets but still within the high humidity of a water cloud environment (see Section 4.1.3). Due to very slow dayside upwelling (on order 0.2 to 0.3 m/s) in the cloud and moderate wind speeds away from the substellar point (<40 m/s within +/_40 degrees longitude), the life particles could spend anywhere from a couple of days to a couple of weeks amidst the cloud. This is ample time for life particles to metabolize, reproduce, and spread. Recall that on Earth the typical generation time of a common sulfate-reducing bacterial anaerobe *Desulfovibrio vulgaris* is between 2 and 3 h [65]. Note that our minimum time estimate is for a particle upwelling through the clouds to a dry, cloud-free altitude layer before it circulates latitudinally or settles back down out of the atmosphere. The longer timescale estimate is for a life particle that manages to circulate within the altitude layer where the clouds exist.

Also worth noting is that particle sizes of 10 micron or less have sedimentation speeds on the order of 0.1 m/s, lower or comparable to the upward equatorial wind on the day side, and lower or comparable to the advective speed, demonstrating that life particles of that size or in droplets of that size can remain aloft against gravitational settling. Particles inside a larger cloud droplet may settle out, as the average droplet size depends on the underlying CCN population mass, and larger particles are heavier with higher sedimentation speeds.

Traveling away from the cloud *en route* to the night side, the life particles and their progeny will dry out as they leave the cloud region. In a dried-out state life would not actively metabolize but nevertheless survive. Downwelling on the night side also has a low velocity of approximately 0.2 m/s, so the journey downwards will take a couple of days.

A key point is that a critical number of life particles must avoid death by downwelling to the too hot layers. Some life particles may re-circulate from the night side back to the planet dayside without reaching destructively hot layers. According to the simulations in [51], atmospheric circulation from the planet night side back to the dayside occurs at a range of altitudes from 0.1 down to 1 bar. At 0.1 bar, the temperatures remain around 320 K and so life particles trapped in this flow zone will survive. However, the temperature gradient is steep between 0.1 and 1 bar and life particles that sink to 1 bar or lower before circulating back to the day side could be at a temperature as high as 500 K, too hot for life of any kind.

We must emphasize a number of caveats to this qualitative picture. First, the K2-18b atmospheric water vapor condenses to ice clouds and not the liquid droplets needed for life. Sub Neptunes slightly cooler than K2-18 b are more likely to have liquid water clouds (see Figure 4) and may even have cooler lower layers where the atmosphere recirculates. Second, the cloud location and extent, and the amount of condensed water is extremely sensitive to a number of factors [51], including particle size (which controls sedimentation rate) and atmospheric metallicity (which sets up dayside temperatures and the dayside radiative timescale, and also the advective timescale by setting up latitudinal temperature gradients). The particle size distribution itself is controlled by the number population of the underlying cloud condensation nuclei (CCN). In addition to the global dayside-to-nightside circulation and particle size, cloud radiative effects and radiative feedback can significantly alter cloud distribution, making it very sensitive to cloud microphysics. As a relevant aside, CCN on sub Neptune exoplanets may be externally delivered in the form of micrometeorites, or may be created in the atmosphere itself from photochemical hazes or condensed salts. For more details, see [51]. Third, despite the dominant dayside/nightside overturning pattern, there are more atmospheric dynamics including jet formation and possible sensitivity of equatorial winds to a number of assumptions that need to be explored in more modeling studies. Finally, while the planet metallicity may be derived or constrained with atmosphere observations, there is no way to determine the CCN number density from either first principles or observations, although cloud variability measured at the planet limb for transiting planets may identify one of the [51] scenarios.

To conclude this section, we point out that some temperate sub Neptunes orbiting M dwarf stars may not yet have completed their dynamical evolution to the tidally-locked state. The planets likely do, however, have slow rotation periods as part of the spin-down evolution. The tidal spin-down time for temperate sub Neptunes is long, because the tidal spin-down equation scales to the sixth power for semi-major axis, and temperate sub Neptunes by definition have moderate semi-major axes. As an example, [51] estimate a tidal spin-down time for K2-18 b ($a$ = 0.14 AU) to be on order 1.7 Ga, assuming the tidal dissipation factor $Q$ is ~$10^4$ (chosen to be inbetween $Q$ values of solar system terrestrial bodies with $Q$ ~ 10 to 500 and giant planets with $Q$ ~ $10^5$ [66]). They [51] compare K2-18 b's tidal spin-down time to the host star K2-18's estimated age of 2.6 ± 0.6 Ga [67]. Given the closeness of the two timescales and considering K2-18b's actual $Q$ value is unknown, and stellar ages are not precisely known, we cannot be certain that K2-18 b or other temperate sub Neptunes are tidally locked, especially cooler planets (i.e., larger semi-major axis) that are more likely to have liquid water condensation in their atmospheres (Figure 4).

The qualitative life particle description in this subsection illustrates what we would like to accomplish in a more quantitative fashion using detailed general circulation model simulations for temperate sub Neptunes that can host liquid water clouds: trace life particles' timescales and pathways. In addition, because of the large parameter space, the community needs a generalized picture for temperate sub Neptunes atmospheric circulation, for example the Weak Temperature Gradient approximation [62] applied to $H_2$ or $H_2$-He-dominated planet atmospheres.

4.1.2. Cloud Location and Downward Transport in Rapid Rotators

We aim to expand our discussion to more rapidly rotating planets than a slowly (and likely synchronously) rotating K2-18b. Another concrete example of atmospheric circulation we can point to is that for Jupiter, a rapidly rotating planet with a 10 h rotation period. While Jupiter is larger and receives less irradiation than a temperate sub Neptune and there remain competing theories for Jupiter's atmosphere banded structure [68], there are some interesting and relevant points about Jupiter's atmospheric flow. We first point out that temperate sub Neptunes orbiting sun-like stars are likely to be rapid rotators as planets are thought to form with high spin rates and only spin down by tidal stresses for planet with orbits close to the host star.

A rapidly rotating planet may have multiple latitudinal "cells" like Jupiter does (e.g., [69,70]). Jupiter's latitudinal cells have fast horizontal winds that rapidly circulate the atmosphere within each cell. The latitudinal cells alternate between westward moving flow (bands) and eastward flow (jets). Relevant here is that Jupiter's latitudinal cells also alternate between upwelling and downwelling air. A sub Neptune atmosphere's latitudinal cells with upwelling air can support water clouds, by replenishing the upper layers with warm moist air from below. See Figure 5b.

Microbial-type life particles can persist aloft in a rapidly rotating planet's atmosphere long enough to reproduce and spread if some of their progeny remain in layers of a latitudinal cell with water clouds. If we think of Jupiter's latitudinal cells, the motion across the cell that leads to downwelling (i.e., along longitudinal lines, called meridional flow) is approximately one order of magnitude slower than the horizontal speeds within a latitudinal cell (called zonal flow). (On Jupiter, the vertical residual wind speeds that are part of the meridional flow are ~1 m/s (e.g., [71]), whereas the horizontal wind speeds are tens of m/s, depending on latitude and altitude (e.g., [69]).) It is, however, worth noting that vertical excursions due to baroclinic eddies could be more important than the weak overturning circulations associated with the midlatitude bands. The baroclinic eddies draw their energy from meridional temperature gradients and may have a significant effect on cloud formation and lifetime, a separate mechanism from Earth's convective cloud formation.

So while many life particles will be carried away from the horizontal flow to the overturning regions, other life particles may remain in the horizontal flow presumably long enough to metabolize and reproduce. Two other key points are (1) the temperature at the bottom of the overturning meridional flow may be too high for life to survive and (2) perhaps the updrafts in the latitudinal cells with upwelling can support life particles against gravitational settling [46,47].

Admittedly, rapidly rotating temperate sub Neptunes are not in the same atmospheric regime as Jupiter, mostly due to the smaller sizes of sub Neptunes (2 to 3) compared to Jupiter (11.2 $R_{Earth}$) but also due to the stronger irradiation of temperate sub Neptunes compared to Jupiter (on order five times greater). We do, however, expect rapidly rotating sub Neptunes to have more bands than the single circulation cell in slow rotators.

What is needed for a general picture of particle flow on planets with different radii, rotation rates, stellar irradiation, and atmospheric mean molecular weight (which affects wave speeds and scale heights) is a general scalable parametrization to capture robust details for atmospheric circulation regimes. The Weak Temperature Gradient approximation [62] is promising in this regard as it represents a key governing factor, the radius of deformation relative to the planet size. For example, temperate sub Neptunes orbiting K stars may have rotation rates in between the slow rotators orbiting M dwarf and rapid rotators orbiting G stars, and consequently a different atmospheric circulation regime than the two categories of planets described above.

4.1.3. Other Physical Factors Affecting Aerial Life

*Life can survive in cold atmosphere regions.* A sub Neptune atmosphere can still be habitable with cold, even seemingly inhospitable top-of-atmosphere layers where life particles might rise up to from lower, warmer atmosphere layers.

Life on Earth does easily adapt to low temperatures. Many organisms on Earth use antifreeze substances to prevent water from freezing and achieve tolerance to extreme cold. For example, synthesis of various cryoprotectants is a common strategy to mitigate extreme cold employed by aerial bacteria in Earth's atmosphere [38]. On Earth, life has been found to metabolize and reproduce at temperatures as cold as at 268 K. Such cold adaptations have evolved independently on many occasions and have been described in virtually all branches of the tree of life—in animals, plants, bacteria, and fungi (see, e.g., [72]). The bacterium *Planococcus halocryophilus* Or1 from high Arctic permafrost actively grows and divides at −15 °C, albeit slowly with one cell division per 25 days, and remains metabolically active at −25 °C [73]. Earth bacteria routinely survive freezing. Some obligatory psychrophiles such as *Colwellia psychrerythraea* 34H are commonly found in sea ice with liquid brine temperatures as low as −35 °C [74], have active motile behavior in temperatures as low as −10 °C [75], and actively reproduce in temperatures as low as −5 °C [76]. Even some complex multicellular organisms (e.g., ice worm *Mesenchytraeus solifugus*) can adapt to spend their entire life cycles in glacial ice and live perpetually below 0 °C [77–79].

*Life can survive in dry atmosphere regions.* Humidity depends on temperature. So, regardless of the atmospheric water content, there can be regions of a sub Neptune atmosphere that are so dry that life particles would dehydrate and have to go into a hibernating phase. For example, below the ice clouds in the K2-18b exoplanet atmospheres with metallicities 100x solar (shown in Figure 4), the relative humidity reaches very low levels, down to 0.3% at a temperature of 390 K and down to approximately 6% at 300 K. For comparison, the driest place on Earth has 2% relative humidity in the Atacama desert (at noon in full sunlight) [80].

On Earth, severe water loss due to even moderately dry air is instantly lethal to most organisms. However, many species of microorganisms, plants, and animals can survive complete desiccation (anhydrobiosis) with no detrimental effects. Moreover, desiccation does not necessarily render active metabolism of microbial cells impossible, as shown for several bacterial species (e.g., [81]). The desiccated organisms can remain in the state of anhydrobiosis for prolonged periods of time without apparent damage and upon rehydration can resume active metabolic processes and reproduction (e.g., [82]). Bacteria and archaea are not the only organisms capable of surviving prolonged periods of dry environmental conditions. Some complex, multicellular organisms (e.g., tardigrades) can remain in suspended animation for several years [83]. Microorganisms also routinely survive dry conditions as spores, in a metabolically inactive state. On Earth, some bacterial spores that are viable can survive in a dormant state in extremely harsh conditions for many thousands [84,85], if not several millions of years [86,87].

*Life could survive inside or outside of cloud droplets.* So far, we have not specified whether life particles must reside inside liquid droplets or whether life particles could metabolize and reproduce outside of liquid droplets while freely floating in the atmosphere. Certainly, the environment inside droplets is most suitable for life because the requirement for a liquid environment is one of the general attributes of all life regardless of its biochemical makeup [56,57]. In addition, the water droplet can protect life from destructive ultraviolet radiation. Most life in Earth's aerial biosphere is inside liquid droplets rather than free floating [37].

The challenge to free-floating life outside of water droplets is the dryness of the atmosphere—life particles would rapidly desiccate (by net loss of liquid to the atmosphere). Specifically, free-floating cells would lose water until their internal water activity is the same as the vapor pressure of the atmosphere around it. On Earth, free-floating metabolically active cells outside of cloud water droplets are known to exist, but they are a small fraction of the overall aerial biomass.

To survive and actively reproduce in atmosphere conditions outside of water droplets, microbial life must have energetically costly active water capture and retention mechanisms. Terrestrial bacteria capture water by using hygroscopic biosurfactant polymers. Many microbial polysaccharides and amphipathic lipopeptides, such as syringafactin, from *Pseudomonas syringae* have highly hygroscopic properties and are instrumental in reducing the water stress of microorganisms [88].

As an aside, life reproducing inside a droplet can populate the atmosphere once the droplet evaporates and the individual life particles become freely floating, perhaps in a dehydrated state. Alternatively, the life particles might develop a mechanism to actively disperse out of the cloud droplets, within the cloud layer. Some Earth ground- based bacteria and microscopic fungi can push out a spore-producing organ through the surface tension of water droplets at the water-air interface over cm to m scales, [89,90] via a relatively complex, energy-intensive mechanism.

*4.2. Active Mechanisms for Complex Life to Persist Aloft*

Active life forms can withstand a planet's atmospheric circulation to purposefully stay aloft and within liquid water cloud regions. Although speculative, it is worth mentioning that complex multicellular life may have several options to actively persist aloft, including active flight and gliding. (Imaginative gas-filled balloons and jet propulsion have also been proposed [46].). On Earth, several species of birds have evolved to remain in the air almost indefinitely either through active flight or gliding. Common swifts (*Apus apus*) have to land only to nest and lay their eggs. Common swifts breed, sleep, collect material for nests, drink rain droplets, and perform all other physiological functions while flying. Many individuals have been recorded flying continuously for ten months [91]. We may speculate that with sufficient evolutionary pressure perpetual active or gliding flight capability is possible on sub Neptunes as well (with vivipary replacing external egg-laying, by analogy with mammals, marsupials, and some species of crocodile, lizard and fish).

## 5. Challenges to Life in a Sub Neptune Atmosphere beyond Persistence Aloft

There are many challenges for any aerial life in an environment without a temperate solid surface beneath. Here, we present the challenges and summarize any logical paths that life might take to overcome them: the origin of life (Section 5.1); nutrient scarcity (Section 5.2); and the issues surrounding UV radiation for sub Neptunes orbiting M dwarf stars (Section 5.3).

*5.1. Origin of Life in an Atmosphere*

The origin of life is possibly the single biggest challenge and perhaps an unsurpassable one to the hypothesis of life existing in a sub Neptune atmosphere. A sub Neptune atmosphere appears highly uniform, in contrast to the diverse environments on a terrestrial planet. Although the exact environment and chemical pathways that led to the formation of life on Earth are unknown, the diversity of environments on Earth is cited as central to the origin of life for one of two reasons: (1) if the correct environment for the origin of life is rare, then having many different environments increases the chances for life to originate, and (2) diverse environments might be required for different parts of the origin of life process (each of which might be rare). It is likely that the environment on early Earth and the prebiotic chemical pathways coevolved to give rise to life. (For more on this extensive topic, see, e.g., [92].) It is unknown whether the cloud environment of sub Neptunes could undergo similar coevolution.

In this subsection, we explain and review key challenges to the origin of life in an aerial environment. We offer possible solutions for some issues.

The origin of life requires concentrated ingredients, a situation at first glance at odds with a massive atmosphere or gas envelope dominated by $H_2$ and He or by $H_2O$. All terrestrial origin of life scenarios envisage environments where key chemicals are generated at high (millimolar to molar) concentrations or concentrated from the environment (as reviewed in [93]). For example, the growing consensus for the origin of life on Earth is for small warm pools of water that cycle between dehydration (to concentrate molecules) and hydration (to enable reactions) (e.g., [94,95])—a situation not present in sub Neptune atmospheres.

Sub Neptunes lack a temperate solid surface environment where complex molecules can readily concentrate. Any solid layer will be far beneath the gas envelope (composed of either $H_2$ or $H_2$ and He, and possibly a large water vapor content) where temperatures are too high for complex molecules. Sub Neptunes and giant exoplanets in general have hot interiors due to residual heat leftover from their formation. Aside from the too-hot temperatures, any sub Neptunes that are water worlds would have a massive high-pressure ice layer, which would likely act as a physical barrier to prevent transport of rocky and metal nutrients up to the envelope and atmosphere. All of this means that life on a sub Neptune would have to originate and survive in the atmosphere.

The origin of life in a sub Neptune atmosphere likely must rely on the delivery of meteoritic material on which the origin of life could happen or for delivery of complex chemicals as feedstock for the origin of life. Delivery of ingredients could come from bombardment by planetesimals shortly after planet formation or from a dynamically unstable asteroid belt, as our solar system has. Several biochemical precursors such as reactive phosphorus species, small molecule metabolites, sugars, amino acids, nucleobases (e.g., adenine), or short chain fatty acids have been found in meteorites [96,97]. It is, however, unknown whether such precursor molecules could subsequently form any complex molecules and molecular systems in a water cloud environment of temperate sub Neptune atmospheres.

Land-based wet–dry cycles on Earth are key in the predominant hypotheses on the environmental conditions required for the origin of life. However, we always keep in mind that there are several alternative (though less popular) hypotheses on where and how life could have originated on Earth. For example, recent work suggests that the early Earth (3–4 Gyr ago) was likely completely covered by a global water ocean [98]. If correct, it becomes difficult to conceive how the classical view of the wet–dry cycles could operate in a global-ocean environment, without dry land.

Here, we emphasize work that shows how cloud droplets may be sites for the origin of life [99] including providing chemical "cycling conditions" not unlike the wet–dry cycles postulated to be crucial in life's origin on Earth's surface. The cloud droplets can provide the necessary concentration and polymerization of prebiotic monomers [100,101]. Water cloud droplets, as they travel through the atmosphere, would be subjected to variable temperature cycles and periodical loss (evaporation) and gain (drop mergers or condensation) of water, a process that could lead to the concentration of chemicals within droplets and therefore facilitation of crucial chemical reactions. One could speculate that CCN could act as a concentrating environment for chemicals by providing a solid substrate (for example, by adsorbing onto frozen ice crystals). A freeze–thaw cycle might further facilitate a type of wet–dry cycle.

In addition to the chemical cycling properties of droplets, several recent studies show that the surface of aqueous droplets provides a special, unique, and generally favorable reaction environment, with qualitatively different thermodynamic and kinetic properties, than bulk aqueous solutions [102,103]. For example, a water micro-droplet environment allows for abiotic production of sugar phosphates and uridine ribonucleoside, both are precursors for nucleic acids [103]. Formation of such compounds is not known to occur in bulk aqueous solutions. Synthesis of peptide bonds (found in proteins on Earth) also in principle could happen at air–water interfaces in atmospheric cloud droplets [104].

Sub Neptune temperate atmospheres likely have the required primary feedstock compounds to satisfy the proposed pathways to generate building blocks of crucial biochemicals such as RNA. Reduced nitrogen-containing compounds are required for this chemistry [105]. It is likely that a reduced atmosphere would contain such gases as $H_2$, $NH_3$, and $CH_4$ [95]. With atmospheres dominated by $H_2$, and temperatures beneath the atmosphere high enough for recycling of molecules to their thermochemical equilibrium forms (for those planets without water oceans), sub Neptune atmospheres are expected to have $NH_3$, as the dominant N-bearing gas, $CH_4$ as the dominant C-bearing gas, as well as $H_2S$ and $PH_3$ (depending on the overall H content of the atmosphere, or "metallicity"). $NH_3$ is a more reactive form of N than $N_2$, aiding chemical processes. From the origin of life perspective, the reducing atmosphere of sub Neptunes not only contains the main elements needed for life (possibly with the exception of metals, see Section 5.2. below) but is indeed more conducive towards prebiotic chemistry than an oxidized atmosphere. A reducing atmosphere allows for the formation of reduced organic molecules, such as nitriles (from photochemistry), and hydrolysis of nitriles can lead to formaldehyde, glycoaldehyde, formamide, and urea, to name a few, that are deemed to be primary and secondary precursors of RNA [95].

In summary, not only might the physical conditions of water clouds appear to be conducive to the origin of life, but the chemical conditions of water cloud droplets might also facilitate critical chemical reactions needed for life to originate, assuming planetesimal or meteorite delivery of other key elements and compounds.

*5.2. Nutrient Scarcity*

A challenge to life's survival in a planetary atmosphere is a lack of nutrients. On Earth, essential non-volatile nutrients (such as metals and mineral salts) reach the atmosphere via uplifted dust and salts from the surface and ocean [38]. There is no rocky layer near a sub Neptune atmosphere (any core is likely $10^3$ to $10^4$ km deep [21]) and therefore no reliable transport of non-volatile nutrients up to the atmosphere. Furthermore, for a water world, a massive interior high-pressure ice layer would likely act as a physical barrier to prevent transport of rocky and metal nutrients up to the envelope and atmosphere.

A major nutrient limitation is lack of metals. Metal ions are ubiquitously required for terrestrial life, where they are used as charge-balancing counter-ions, as key ligands conferring protein specificity, as activating groups in catalysis and as enablers of a wide range of redox chemistry. More than one-third of all proteins in terrestrial eukaryotes and bacteria require metal binding to function properly [106]. Note that the elements CHNOPS are commonly available, based on solar abundances and their presence in common volatile molecules.

Iron could be delivered to a sub Neptune planet atmosphere by meteoritic delivery. The current accretion rate of meteoritic material to the Earth is of the order of 20–70 kilotons year$^{-1}$ [107]. Approximately 6% of this delivered material is in the form of iron/nickel meteorites [108]. Under the assumption that Fe-Ni meteorites consist of 95% iron, the maximal meteoritic delivery of Fe to Earth is in the range of $4.0 \times 10^9$ g year$^{-1}$. For context, global Fe assimilated by oligotrophic phytoplankton in the open oceans reaches $6.7 \times 10^{11}$ g year$^{-1}$ [109], approximately two orders of magnitude higher than the total amount of Fe reaching Earth through meteoritic delivery. However, the meteoritic delivery rate is certainly enough to support a modest biomass in an aerial biosphere. A sub Neptune would need to be a part of a planetary system with an unstable asteroid belt in order to have a steady flux of meteoritic material over billions of years. At present, we are unable to detect the presence of exoplanet asteroid belts, though next-generation telescopes might be able to [2–7].

The situation for other metals including Mg, Ca, Cu, Zn, Co, and K is bleaker. The estimated global amounts of metals assimilated by phytoplankton in Earth's oceans are five to nine orders of magnitude higher than our estimate for meteoritic delivery of these metals to Earth (following the above calculation for Fe and using values from [60]). We note that such a comparison does not take into account any adaptations that life might develop to mitigate nutrient scarcity. Life on Earth inhabits environments where nutrients are extremely scarce with adaptations such as: storage of materials, as a source of elements in an event of extreme shortages; recycling and reuse of already acquired nutrients; and metabolic flexibility. Such adaptations are present in microbial communities in the lower oceanic crust [110] or in Earth's aerial biosphere where life has developed a series of specific adaptations for efficient capture of limiting nutrients, including siderophore-mediated transition metal capture [38]. It is conceivable that life in any sub Neptune aerial biosphere, if it exists, has evolved similar solutions for efficient scavenging and recycling of metallic trace elements. Some of the roles played by metals in Earth life could be played by salts such as $NH_4Cl$, which, for example, should form near 0.1 bar for K2-18b's simulated conditions [51].

Nutrient and metal delivery could come from an orbiting exomoon. Due to tidal forcing from the host planet, an exomoon may send several orders of magnitude more material to the host planet than the amount of material delivered to Earth by meteorites [111]. For comparison's sake, the close-in exomoons studied around giant exoplanets would provide ~31 gigatons yr$^{-1}$. Additionally, two relevant solar system examples are that Jupiter's tidally-heated moon Io vents ~ 1 ton s$^{-1}$ into space (31 megatons yr$^{-1}$) [112] and the Pluto-Charon system exhibits mass transfer [113]. We note, however, the true degree to which mass transfer from a moon to a planet is possible and hence an efficient source of nutrients is unknown and requires further study. While exomoons have not yet been discovered around exoplanets, their detection is feasible and such exomoons may also be habitable [114]. If such exomoons are inhabited, the clouds of a temperate sub Neptune could be seeded with life through panspermia, perhaps continuously.

Without an unstable asteroid belt and meteorite delivery and efficient strategies for recycling and reusing metals, a sub Neptune aerial biosphere might have to function without metals. Could life exist without metals entirely? This is a very speculative subject. For terrestrial biochemistry the answer is definitely 'no', but the broad classes of functions summarized above can all be carried out by non-metal compounds. Charge balancing can be done with ammonium compounds ($NH_3^+$ groups) or carboxylic acids [115]. The many enzymes that work without a metal ion in their active site demonstrate that metals are not universally essential for catalysis. Additionally, redox chemistry, even single-electron redox chemistry, can be done with organic compounds, especially sulfur compounds and derivatives of nitrogen oxides. Thus, we cannot logically conclude that metals are absolutely essential for life, only that on Earth, where metals are available in the crust, they are widely and universally used. We also note that many scenarios for the origin of life depend on metals and minerals, so metals might be essential for life to arise in the first place.

*5.3. High-Energy Radiation or Lack Thereof for Planets Orbiting M Dwarf Stars*

A challenge for the origin of life on planets orbiting M dwarf stars is the quiescent phase deficiency of UV radiation needed for generation of precursor RNA molecules. M dwarf stars have 10–1000 less UV radiation than sun-like stars [116]. It has been proposed, however, that the high-energy radiation, including UV, coming during frequent flares from M dwarf stars [117,118] can substitute for the continuous UV radiation the sun outputs. However, there is not yet a laboratory demonstration that supports the idea that intensive bursts (with duration of minutes to hours) every 10 days or so is a suitable replacement for continuous UV radiation. Note that only two stars have had UV measured during a flare ([119,120]).

Near-UV radiation in the range 200 nm < $\lambda$ < 280 nm may be a key factor behind many prebiotic processes such as ribonucleotide and sugar synthesis pathways [117,121]. According to the current leading origin of life paradigm, the synthesis of RNA precursors may rely on a critical photochemical step involving NUV radiation at 254 nm. Recent work quantifies the radiation fluxes needed for such synthesis ($2 \times 10^9$ to $10^{10}$ photons cm$^{-2}$ s$^{-1}$ A$^{-1}$) and argues that while M dwarf stars lack the required UV flux, flaring activity on approximately 30% of stars cooler than stellar types K5, and ~ 20% of stars of early M dwarf type is adequate to reach the required levels at the surface of early Earth-type planet atmospheres [117]. We note that more recent data from the MIT-led NASA TESS mission [118] can be used to update these values and extend them to cooler M dwarf stars.

The stellar NUV radiation in the range 200–300 nm can penetrate a planet atmosphere to at least one bar [116]. Life in the clouds of temperate sub Neptunes may be at that altitude or above, due to the high temperatures in the lower atmosphere (Figure 4 and references therein). Relevant for any reduced atmosphere is that $H_2O$ absorbs at wavelengths >300 nm. No dominant atmosphere species expected in temperate sub Neptune atmospheres absorbs in the 200–300 nm range ($CH_4$, $PH_3$, and CO or $CO_2$ if present) [6]. One exception is $H_2S$, but because it absorbs at wavelengths <230 nm $H_2S$ does not interfere with the proposed set of reactions involving 254 nm radiation. As well, $H_2S$ is not expected to be very abundant. While $O_3$ absorbs blueward of 300 nm, its existence requires the presence of $O_2$.

Some argue that X-ray, EUV radiation, and high-energy particles coming from flares might be a challenge for a biosphere due to harmful effects on life or even by heavily eroding the planet atmosphere. During a flare event, some atmospheric life particles would be protected if they were on the night side, although life particle residence times must be considered. Some life on Earth can resist high-energy radiation, although such species are relatively few and are highly adapted to the specific environment [122,123]. We know that sub Neptune planets do not lose all of their atmospheres, even those around active M dwarf stars, based on their low average densities that require existence of an atmosphere. Indeed, many sub Neptune atmospheres may survive indefinitely: the population of sub Neptunes orbiting old and quiet M dwarfs have survived any flares and high-energy output for many Myr to Gyr, including the time period when their host stars were young and most active. This is under the accepted stellar evolution theory that M dwarf stars are born as fast rotators (i.e., highly active) and then slow down as they grow older to a quiescent phase (e.g., [124,125]). That said, many smaller exoplanets with higher average densities than sub Neptunes are deemed to be end products of severe atmosphere erosion, even if they retain thin atmospheres [126–128].

## 6. Possible Metabolic Strategies and Biosignatures

All life needs a metabolic strategy to gain energy from its environment. In the search for life beyond Earth we make an underlying assumption that life uses chemistry to extract energy from its environment, to store energy, and to use energy. In the process life will output a byproduct waste gas or gases, some of which can accumulate in an atmosphere to be considered biosignature gases. Life in an aerial biosphere may in principle produce many of the biosignature gases considered elsewhere (for a review of biosignature gases, see [129,130]).

As a kind of "proof of concept" we show two possible metabolic strategies for life in a sub Neptune aerial biosphere: hydrogen photosynthesis (Section 6.1) and methanogenesis (Section 6.2). We also describe life's light-harvesting pigments as potential biosignatures (Section 6.3).

*6.1. Hydrogenic Photosynthesis and CO as a Bioindicator*

The conversion of light energy to chemical energy through photosynthesis is one of the oldest and most fundamental energy-harnessing biochemical processes on Earth, suggesting it will be a fundamental strategy on other planets. There are many types of photosynthesis that life on Earth uses for carbon fixation, beyond the most common oxygenic photosynthesis that can be summarized as $CO_2 + H_2O \rightarrow CH_2O + O_2$ reaction (others reviewed in [131]). Many more types of photosynthesis are theoretically possible [132–134]. Photosynthesis need not be oxygenic, or even for carbon fixation. Some life on Earth uses light energy to run chemical reactions and to generate and store energy for reasons not limited to carbon fixation [135].

Here, we summarize "hydrogenic photosynthesis" [134], with a net reaction

$$CH_4 + H_2O \rightarrow CH_2O + 2H_2.$$

The substrates $CH_4$ and $H_2O$ for hydrogenic photosynthesis are readily available in the sub Neptune atmosphere, as $CH_4$ is expected to be the dominant form of carbon and $H_2O$ the dominant form of oxygen [136].

The hydrogenic photosynthesis product, $CH_2O$, represents organic molecules containing a carbonyl group (C = O) such as aldehydes or ketones, including sugars that are used to build life forms. Organic molecules containing a carbonyl-group could include small volatile aldehydes and ketones, including formaldehyde.

In separate work, we have found the $CH_2O$ category of volatile molecules are very poor biosignature gases because their high reactivity means that they cannot accumulate in any atmosphere (Zhan et al. in prep.). However, the end reactivity product of volatile $CH_2O$-containing molecules is CO, which may be a bioindicator in $H_2$-dominated atmospheres that lack $CO_2$, if as high amount of volatile $CH_2O$ is produced as $O_2$ on Earth (Zhan et al., in prep). Furthermore, the host star must be an M dwarf star with far-UV (~121–200 nm) > NUV (~200–300 nm) to enable CO to accumulate instead of $CO_2$ against photochemical reactions.

*6.2. Methanogenesis*

Methanogenesis is a process where life catalyzes the reduction of $CO_2$ to $CH_4$ to release energy. The reduction of $CO_2$ by $H_2$ would represent a ubiquitous source of energy for life on any world with an $H_2$-rich atmosphere, including sub Neptunes.

Methanogenesis can proceed even under conditions where only trace amounts of $CO_2$ are available. The reduction of $CO_2$ by $H_2$ in water yields 10 kJ mol$^{-1}$, under an atmosphere containing $10^{-72}$ as little $CO_2$ as $CH_4$ and 70% $H_2$ at 1 bar [134]. Such energy yield is the minimum free energy usable by terrestrial methanogens [137]. Thus, life can yield enough energy released by the reduction of $CO_2$ to $CH_4$, with small amounts of $CO_2$ being regenerated by photochemistry.

The methanogenesis reaction proceeds in an aqueous environment as follows:

$$CO_2(aq) + 4\,H_2(aq) \rightarrow CH_4(aq) + 2\,H_2O(l); \Delta G^\circ = -193\text{ kJ/mol at }25\,^\circ C.$$

At the life-supporting temperatures of sub Neptune cloud layers, we expect some $CO_2$ to be available. Although the dominant form of carbon in a $H_2$-dominated sub Neptune atmosphere is $CH_4$, there will always be some CO and a small amount of $CO_2$ [136,138]. In particular $CO_2$ can be produced at high altitudes from the photochemistry of CO and $H_2O$.

The metabolic byproduct gases $CH_4$ and $H_2O$ are not useful biosignature gases for a planetary atmosphere that already has $CH_4$ and $H_2O$ as the dominant form of carbon and oxygen, respectively. That said, it is worth investigating the isotopic observability of $CH_4$ at different wavelengths with future telescopes because biogenic methane is likely to have a different isotopic signature from $CH_4$ produced abiologically from the $H_2$ and carbon species available. For feasibility simulations of carbon isotope detections with extremely large telescopes now under construction, see [139].

We conclude our discussion of hydrogenic photosynthesis and methanogenesis by emphasizing their complementarity. The product of methanogenesis ($CH_4$) could be the substrate for the hydrogenic photosynthetic carbon fixation in situations where needed.

*6.3. Light-Harvesting Pigments*

Light-harvesting organisms use pigments to absorb light energy. Chlorophyll, for example, is used to harvest sunlight by plants and photosynthetic microorganisms on Earth. Life on Earth uses a range of pigments that, though less widespread than chlorophyll, populate most of the visible wavelength range, showing the potential diversity of light-harvesting pigments. Pigments are in fact produced in very high abundance and diversity by life on Earth, including melanins, carotenoids and other pigments, for protection against sunlight, but also for many other reasons.

For background, astronomers have toyed with the idea that light-harvesting pigments may be good biosignatures. Earth's vegetation's "red edge" motivated a flurry of activity the early 2000s [140–144] based on the phenomenal property that some plant leaves are over 25 times more reflective at near-IR wavelengths than at visible wavelengths. This vegetation "red edge" signature is extensively used in satellite monitoring for the health of Earth's forests and plant crops. That said, Earth's extensive vegetation cover is not enough for the red edge to be a strong biosignature for the test case of Earth as an exoplanet because large tracts of forest are usually cloud covered and only part of Earth is visible in reflected light at any illumination phase [142]. Nonetheless, researchers have postulated pigments of various kinds have a chance to be a strong biosignature on exoplanets (e.g., [142,145,146]).

We can, nevertheless, keep open to the idea that any microbial life in a sub Neptune aerial biosphere might also have an observable UV or visible-light pigment.

**7. Target List of Temperate Sub Neptunes Suitable for Atmospheric Observations**

There is a growing list of temperate sub Neptune exoplanets to explore for habitability. Sub Neptunes appear to be the most common type of planet in our Galaxy [16–18], as far as discovery selection effects allow us to determine. Furthermore, sub Neptune planets may have temperate regions in their atmospheres where water can exist in a liquid form even for planets beyond the classical habitable zone.

The ultimate goal is to observe the planet's atmosphere and infer the presence of water clouds with liquid droplets and to search for biosignature gases. To evaluate a sub Neptune's suitability for atmosphere measurements, we adopt the transmission spectrum detection metric from [147]. Together with the planet's atmospheric temperature, and considering planets with radii > 1.5 $R_{Earth}$, the transmission spectroscopy metric is our main criteria for highlighting target sub Neptune exoplanets for future observation.

*7.1. Temperature and Scaled Semi-Major Axis Suitability Metric*

The atmospheric temperature is the key parameter for a sub Neptune to host an aerial biosphere. Somewhere in an atmospheric layer the temperature must lie in a suitable range for life. Life on Earth exists in the range of 258–395 K. Above 395 K temperatures are so high that the great majority of proteins denature. While water freezes out below 258 K, life is able to adapt to colder temperatures (see Section 4.1.3).

The atmospheric layer with the right temperature also has to be at a suitable pressure for liquid water. For example, for 394 K water can be liquid at pressures only higher than 5 bar, that is five times Earth's surface pressure. Moreover, the pressure at which water is liquid rather than ice also depends on the atmospheric water content (see Section 3).

Taking the above considerations for life's and liquid water's temperature range, for our planet target search we adopt an atmosphere equilibrium temperature of 200 K $\leq T_{eq} \leq$ 320 K, using the water phase diagram in Figures 3 and 4 as a guide.

The planet equilibrium temperature $T_{eq}$ is a representative temperature of the planet atmosphere, typically corresponding to a layer near the top of the atmosphere. For background, $T_{eq}$ is the effective temperature attained by an isothermal planet after it has reached complete equilibrium with the radiation from its host star. Although planet atmospheres are not isothermal, temperatures in the upper atmosphere can be similar across a few orders of magnitude in pressure, making $T_{\rm eq}$ an adequate approximation.

$T_{\rm eq}$ is not directly measurable but connected to the planet semi-major axis (itself derived from observations) via

$$T_{\rm eq} = T_* \sqrt{R_*/a} [f(1 - A_B)]^{1/4}, \tag{1}$$

for a planet in a circular orbit. Here, $T_*$ is the star's effective temperature, $R_*$ is the star's radius, and $A_B$ is the planet's Bond albedo. $f$ is a correction factor that equals ¼ if the absorbed stellar radiation is uniformly redistributed around the planet, that is, into 4π. For a slowly tidally locked or very slowly rotating planet, the stellar energy is absorbed only by one hemisphere of the planet. If the atmosphere instantaneously reradiates the absorbed radiation (with no advection), $f$ = ⅔ (e.g., [148]).

To select target planets, we have to use a planet's semi-major axis via Equation (1), because in advance of atmosphere measurements, the planet atmosphere temperature is not known. Even with observed spectra, it is challenging to accurately extract a temperature–pressure profile because most retrieval models are 1D and cannot take into account tidal locking and 3D dynamical atmospheres. Furthermore, transmission spectra are not very sensitive to temperature.

*7.2. Target Star List*

We have created a list of currently known and candidate sub Neptunes suitable for atmospheric investigation for liquid water clouds based on a planet's radius, $T_{eq}$, and transmission spectra metric (TSM) [147]. Our final criteria are:

- $R_{\rm p}$ > 1.5 $R_{\rm Earth}$,
- 200 K ≤ $T_{\rm eq}$ < 320 K, and
- TSM > 12.

The radius cutoff is based on empirical data that show planets with $R_{\rm p}$ > 1.5 $R_{\rm Earth}$ or > 1.6 $R_{\rm Earth}$ are not predominantly rocky [20,149] but must have a low-density component such as an $H_2$/He envelope. Our list may contain a few super Earth "contaminants", because some planets with small radii are on the boundary of super Earth and sub Neptune planets such that future atmosphere observations are needed to definitively categorize them. We therefore name a special category "tiny sub Neptunes". We adopt the following planet categories:

- $R_{\rm p}$ ≤ 1.5 $R_{\rm Earth}$ "Earths and super Earths";
- 1.5 $R_{\rm Earth}$ < $R_{\rm p}$ ≤ 1.8 $R_{\rm Earth}$ "tiny sub Neptunes";
- 1.8 $R_{\rm Earth}$ < $R_{\rm p}$ ≤ 2.75 $R_{\rm Earth}$ "small sub Neptunes";
- 2.75 $R_{\rm Earth}$ < $R_{\rm p}$ ≤ 4 $R_{\rm Earth}$ "large sub Neptunes".

While we choose the criteria TSM > 12 (blue-shaded region in Figure 6) the cutoff is merely a representation of the best set of temperate sub Neptunes for atmospheric follow up with the JWST.

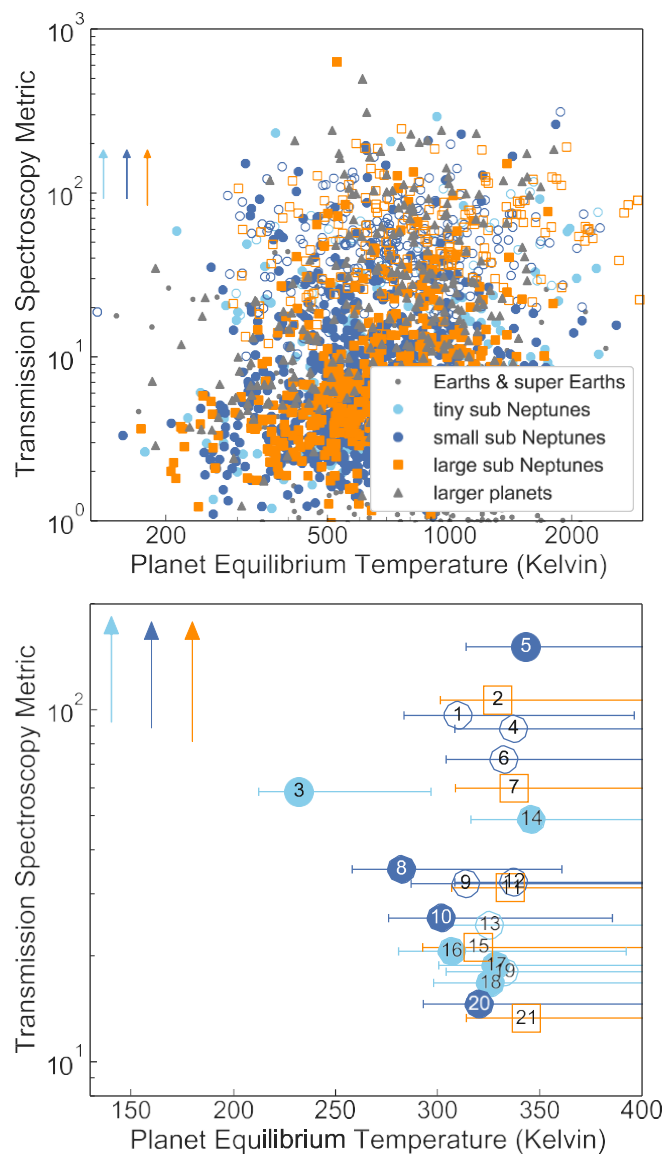

**Figure 6.** Temperate sub Neptune exoplanets and planet candidates suitable for hosting an aerial biosphere (blue-shaded region). The atmospheric transmission spectroscopy metric (TSM; $y$ axis) vs. planet equilibrium temperature in Kelvin ($T_{eq}$; $x$ axis). Planet categories are indicated as follows: grey circles for Earths and super Earths (<1.5 $R_{Earth}$); light blue circles for tiny sub Neptunes (1.5 to 1.8 $R_{Earth}$); blue circles for small sub Neptunes (1.8 to 2.75 $R_{Earth}$); orange squares for large sub Neptunes (2.75 to 4 $R_{Earth}$); and grey triangles for planets larger than sub Neptunes (>4 $R_{Earth}$). Confirmed sub Neptunes have filled symbols, while planet candidates (TESS Objects of Interest, TOIs) have unfilled symbols. Central values stem from assuming zero Bond albedo and full heat-recirculation (see Equation (1)). Error bars reflect different Bond albedo and heat-recirculation assumptions (Equation (1); with Bond albedo, $A_B$, of 0 or 0.3, and correction factor, $f$, of ¼ or ⅔.). The three arrows show the recommendation for a minimum TSM [147] color coded for different planet sizes (TSM > 92 for 1.5 < $R_p$ < 2.75 and TSM > 84 for $R_p$ > 2.75). The blue-shaded region (TSM > 10 and $T_{eq}$ < 320 K) indicates the favorable sub Neptune parameter space for liquid water clouds, in terms of observability and cold enough temperatures. The prime candidates falling into this region include nine confirmed planets LHS 1140 b, LP 791-18 c, K2-18 b, K2-9 b, TOI-1266 c, K2-3 d, Kepler-560 b, K2-133 e, K2-125 b as well as 12 candidate TOIs (see Figure 7). All data shown rely on estimated equilibrium temperatures, estimated transmission spectroscopy metrics, and some planets and candidates rely on estimated planet masses, described in the main text. The right-side panel is a zoom in of the left-side panel. The numbers indicate a ranking described in the Figure 7 caption. Data taken from the NASA Exoplanet Archive and the TOI Catalog on 28 January 2021.

| Ranking | Planet Name | TIC ID | Score | TSM | Teq (K) (AB=0 f=1/4) | Teq (K) (AB=0.3 f=1/4) | Teq (K) (AB=0f =2/3) | Rstar (RSun) | Mstar (Msun) | Teq star (K) | Rp (REarth) | Period (d) |
|---|---|---|---|---|---|---|---|---|---|---|---|---|
| 1 | TOI-1955.01 (candidate) | TIC 166184426 | 96.48 | 96 | 310 | 284 | 396 | 0.31 | 0.29 | 3327 | 2.23 | 16.32 |
| 2 | TOI-1231.01 (candidate) | TIC 447061717 | 70.68 | 107 | 329 | 301 | 421 | 0.47 | 0.46 | 3567 | 3.28 | 24.25 |
| 3 | LHS 1140 b | TIC 92226327 | 34.08 | 59 | 232 | 212 | 297 | 0.21 | 0.18 | 3216 | 1.73 | 24.74 |
| 4 | TOI-1468.01 (candidate) | TIC 243185500 | 33.88 | 88 | 337 | 308 | 431 | 0.37 | 0.36 | 3382 | 2.33 | 15.53 |
| 5 | LP 791-18 c | TIC 181804752 | 31.06 | 151 | 343 | 314 | 439 | 0.17 | 0.14 | 2960 | 2.31 | 4.99 |
| 6 | TOI-1452.01 (candidate) | TIC 420112589 | 30.53 | 72 | 332 | 304 | 425 | 0.28 | 0.25 | 3248 | 2.01 | 11.06 |
| 7 | TOI-1806.01 (candidate) | TIC 166648874 | 15.65 | 60 | 337 | 309 | 431 | 0.40 | 0.39 | 3272 | 3.41 | 15.15 |
| 8 | K2-18 b | TIC 388804061 | 10.23 | 35 | 282 | 258 | 361 | 0.41 | 0.36 | 3457 | 2.37 | 32.94 |
| 9 | TOI-2285.01 (candidate) | TIC 329148988 | 8.04 | 32 | 314 | 287 | 401 | 0.46 | 0.46 | 3546 | 1.90 | 27.27 |
| 10 | K2-9 b | TIC 82050863 | 4.36 | 26 | 302 | 276 | 386 | 0.31 | 0.3 | 3390 | 2.25 | 18.45 |
| 11 | TOI-876.01 (candidate) | TIC 32497972 | 3.81 | 31 | 336 | 307 | 429 | 0.60 | 0.59 | 3882 | 3.15 | 38.7 |
| 12 | TOI-782.01 (candidate) | TIC 429358906 | 3.32 | 32 | 337 | 308 | 431 | 0.41 | 0.4 | 3320 | 2.00 | 16.05 |
| 13 | TOI-2094.01 (candidate) | TIC 356016119 | 3.06 | 24 | 325 | 298 | 416 | 0.38 | 0.37 | 3457 | 1.54 | 18.79 |
| 14 | TOI-1266 c | TIC 467179528 | 2.77 | 49 | 346 | 316 | 442 | 0.42 | 0.45 | 3600 | 1.56 | 18.80 |
| 15 | TOI-2088.01 (candidate) | TIC 441765914 | 2.68 | 21 | 320 | 293 | 409 | 0.83 | 0.80 | 4902 | 3.58 | 124.73 |
| 16 | K2-3 d | TIC 173103335 | 2.21 | 21 | 307 | 281 | 392 | 0.56 | 0.6 | 3896 | 1.53 | 44.56 |
| 17 | Kepler-560 b | TIC 239275865 | 1.11 | 19 | 329 | 301 | 420 | 0.33 | 0.34 | 3556 | 1.72 | 18.48 |
| 18 | K2-133 e | TIC 150096001 | 0.79 | 17 | 326 | 298 | 416 | 0.46 | 0.46 | 3655 | 1.73 | 26.58 |
| 19 | TOI-2276.01 (candidate) | TIC 198456933 | 0.75 | 18 | 333 | 304 | 425 | 0.88 | 0.96 | 5482 | 1.78 | 153.72 |
| 20 | K2-125 b | TIC 405673618 | 0.46 | 15 | 320 | 293 | 409 | 0.4 | 0.49 | 3654 | 2.18 | 21.75 |
| 21 | TOI-2294.01 (candidate) | TIC 284361752 | 0.05 | 13 | 343 | 314 | 439 | 1.28 | 1.29 | 6424 | 3.50 | 340.38 |

**Figure 7.** Temperate sub Neptune exoplanets suitable for atmosphere follow up. The planets are ranked by their position with respect to favorable atmospheric temperature and the Transmission Spectroscopy Measurement (TSM) (the blue region in Figure 6). A target with a median value fully within the favorable parameter range has a score = TSM. If the target is outside the favorable range, the target's score gets down-weighted according to how far the target's parameter's median lies outside the favorable range. Quantitatively, score = TSM * f(Teq) * g(TSM). Here, the term f($T_{eq}$) = (320 K − $T_{\text{eq\_low}}$)/($T_{\text{eq}}$ − $T_{\text{eq\_low}}$) for $T_{\text{eq}}$ < 320 K. The term g(TSM) = (TSM − 12)/(x − 12) if TSM < x, 92 with x = 92 for tiny and small sub Neptunes (1.5 $R_{\text{Earth}}$ < $R_P$ ≤ 2.75 $R_{\text{Earth}}$), and x = 84 for large sub Neptunes 2.75 $R_{\text{Earth}}$ < $R_P$ ≤ 4 $R_{\text{Earth}}$ [147].

We retrieved all confirmed exoplanets from the Planetary Systems Composite Data table provided by the NASA Exoplanet Archive and a list of all current TESS Objects of Interests (TOIs) from the TESS website (on 28 January 2021) (Figure 6). We only regard TOIs which are considered valuable planet candidates by the TESS follow up team (TFOP) by their TFOP master priority of 1 or 2 (1: high, 5: low). We also estimate the planet masses, if not available, using the approach by [150]. Then, we estimate the TSM [149].

The most favorable targets are ranked according to their median values' proximity to the blue-shaded region (Figures 6 and 7). The prime candidates include the six confirmed planets LHS 1140 b, LP 791-18 c, K2-18 b, K2-9 b, TOI-1266 c, K2-3 d, K2-133 e, and K2-125 b, as well as 12 candidate planet TOIs (see Figure 7), although some of these fall into the tiny sub Neptune category and may be super Earths.

The TESS mission [26] allows the exploration of a whole new region of sub Neptune parameter space as the TESS extended mission is sensitive to temperate sub Neptunes transiting relatively bright M dwarf stars.

## 8. Summary and Future Work

It is natural to wonder whether temperate sub Neptune exoplanets may have an aerial biosphere. We were motivated by the fact that sub Neptunes have extensive "puffy" atmospheres, making them amenable to observational study in the near-term future, possibly even giving a short-cut for biosignature gas searches. Unlike larger, giant exoplanets, sub Neptunes are far more common in the temperate temperature range for orbital periods < 100 days, meaning plenty of favorable planets transiting M dwarf stars will be available for atmospheric study. Temperate sub Neptunes are a unique population, and deserve broad and deep attention.

The key work needed to further the idea of an aerial biosphere on a temperate sub Neptune is atmospheric circulation models for liquid water cloud formation and distribution. This modeling is faced with a huge range of unknown yet controlling input parameters including cloud condensation nuclei [51] and interior heat flow. In addition, computation of Lagrangian trajectories based on both large-scale 3D winds and settling are needed to follow microbial-type life particles to understand whether a critical mass of life can survive aloft. This includes the residence time of life particles amidst liquid water clouds (needed for life to metabolize and reproduce), and whether a critical mass of life particles can avoid downwelling to destructively high temperatures. Recall that without a temperate surface as a barrier to descent, a descending organism may reach an environment where the planet's temperatures will be too hot for life of any kind to survive, because planet atmospheres ultimately get warmer with decreasing altitude.

A sample of sub Neptune-sized planets cooler than the iconic K2-18b will aid in the search. Such planets may have cool enough temperatures (<395 K) for life to survive at the low altitudes where the slowly rotating planet atmosphere recirculates from the night side to the day side. In addition, K2-18b's atmospheric water vapor can condense only to ice and not liquid needed for an aerial biosphere, and here cooler planets would help as well (Figure 4). While we focused on sub Neptunes orbiting M dwarf stars, the prospects and challenges for aerial life are equally relevant to exoplanets with sun-like host stars. Atmospheres of temperate sub Neptunes orbiting sun-like stars can be observed much more easily than the atmospheres of smaller Earth-sized exoplanets, hence we should keep the possibility of sub Neptune aerial biospheres in mind when high-contrast space-based direct imaging missions become a reality in the coming decades.

The most severe and limiting challenge to life in a sub Neptune atmosphere is the origin of life. While we reviewed a few arguments for cloud particles' possible wet–dry cycles to concentrate ingredients, these are by no means proven. However, it is fair to say we do not know where life originated on Earth or even if such wet–dry cycles were indeed crucial.

Nutrient scarcity is the second most severe challenge to life in a sub Neptune atmosphere. While a planet atmosphere might receive enough iron from an unstable asteroid belt if the planet receives meteorites of similar types and rates to what Earth does, all other metals would not be delivered in high enough quantities by orders of magnitude. Life would therefore require strategies to recycle and reuse metals, or to replace metal chemistry with non-metallic chemistry. Note that there is no conventional surface on sub Neptunes from which minerals can be upswept into the aerial biosphere. Any solid surface of ice or an ice/rock mixture would be deep and hot and unable to concentrate chemicals needed for the origin of life.

We conclude with a scenario where life could exist and thrive in a temperate sub Neptune atmosphere, albeit one with much lower biomass than in Earth's oceans. Imagine a situation where a terrestrial planet hosts life but that planet cannot be observed (non-transiting, or too small, for example). The terrestrial planet could be in a system with a temperate sub Neptune that is more orbitally distant from the host star (favorable for the cooler lower atmosphere layers, reasons given above) and an unstable asteroid belt. If interplanetary transfer of impact ejecta from the terrestrial planet and containing living material ("panspermia") could seed life in the sub Neptune's liquid water clouds, the severe challenges of the origin of life would be eliminated. The unstable asteroid belt would serve to deliver meteoritic material containing nutrients to the sub Neptune atmosphere. An additional scenario is the existence of an undetectable (and possibly inhabited) exomoon orbiting very close to the sub Neptune (in place of an inhabited terrestrial planet). Tidal forces from the planet could induce volcanic activity that leads to mass transfer of nutrients [111] to the sub Neptune atmosphere.

In conclusion, while there are many challenges and unknowns for the conditions to support an aerial biosphere on temperate sub Neptunes, we can keep an open mind to be aware of any unusual gases that might be biosignatures observed with upcoming, next-generation telescopes.

**Author Contributions:** Conceptualization, S.S., J.J.P. and W.B.; investigation, S.S., J.J.P. and M.N.G.; resources, S.S; data curation, M.N.G.; writing—original draft preparation, S.S; review and editing, S.S., J.J.P., W.B., M.N.G., T.M.-E., D.D.; visualization, S.S., J.J.P. and M.N.G.; project administration, S.S.; funding acquisition, S.S. All authors have read and agreed to the published version of the manuscript.

**Funding:** This research was funded by in part by NASA, grant number 80NSSC19K0471, the Heising-Simons Foundation, grant number 2018–1104, and the MIT Torres Fellowship program (M.N.G.).

**Institutional Review Board Statement:** Not applicable.

**Informed Consent Statement:** Not applicable.

**Data Availability Statement:** Not applicable.

**Acknowledgments:** We thank Ray Pierrehumbert for extensive discussions which improved Section 4 of this paper. We thank Benjamin Charnay, Rick Binzel, Paul Rimmer, and Ewa Borowska for useful discussions. We thank Joanna Petkowska for preparation of Figure 5.

**Conflicts of Interest:** The authors declare no conflict of interest. The funders had no role in the design of the study; in the collection, analyses, or interpretation of data; in the writing of the manuscript, or in the decision to publish the results.

## References

1. Gardner, J.P.; Mather, J.C.; Clampin, M.; Doyon, R.; Greenhouse, M.A.; Hammel, H.B.; Hutchings, J.B.; Jakobsen, P.; Lilly, S.J.; Long, K.S.; et al. The James Webb Space Telescope. *Space Sci. Rev.* **2006**, *123*, 485–606. [CrossRef]
2. Gilmozzi, R.; Spyromilio, J. The European Extremely Large Telescope (E-ELT). *Messenger* **2007**, *127*, 11.
3. Johns, M.; McCarthy, P.; Raybould, K.; Bouchez, A.; Farahani, A.; Filgueira, J.; Jacoby, G.; Shectman, S.; Sheehan, M. Giant magellan telescope: Overview. *Int. Soc. Opt. Photonics* **2012**, *8444*, 84441H.
4. Sanders, G.H. The Thirty Meter Telescope (TMT): An International Observatory. *J. Astrophys. Astron.* **2013**, *34*, 81–86. [CrossRef]
5. Seager, S.; Turnbull, M.; Sparks, W.; Thomson, M.; Shaklan, S.B.; Roberge, A.; Kuchner, M.; Kasdin, N.J.; Domagal-Goldman, S.; Cash, W.; et al. The Exo-S probe class starshade mission. *Tech. Instrum. Detect. Exopl. Vii* **2015**, *9605*, 96050W. [CrossRef]
6. Gaudi, B.S.; Seager, S.; Mennesson, B.; Kiessling, A.; Warfield, K.; Cahoy, K.; Clarke, J.T.; Domagal-Goldman, S.; Feinberg, L.; Guyon, O.; et al. The Habitable Exoplanet Observatory (HabEx) Mission Concept Study Final Report. *arXiv* **2020**, arXiv:2001.06683.
7. The LUVOIR Team The LUVOIR Mission Concept Study Final Report. *arXiv* **2019**, arXiv:1912.06219.
8. Meixner, M.; Cooray, A.; Leisawitz, D.; Staguhn, J.; Armus, L.; Battersby, C.; Bauer, J.; Bergin, E.; Bradford, C.M.; Ennico-Smith, K.; et al. Origins Space Telescope Mission Concept Study Report. *arXiv* **2019**, arXiv:1912.06213.
9. Adams, E.R.; Seager, S.; Elkins-Tanton, L. Ocean planet or thick atmosphere: On the mass-radius relationship for solid exoplanets with massive atmospheres. *Astrophys. J.* **2008**, *673*. [CrossRef]
10. Lopez, E.D.; Fortney, J.J. Understanding the Mass-Radius Relation for Sub-neptunes: Radius as a Proxy for Composition. *Astrophys. J.* **2014**, *792*, 1. [CrossRef]
11. Montet, B.T.; Morton, T.D.; Foreman-Mackey, D.; Johnson, J.A.; Hogg, D.W.; Bowler, B.P.; Latham, D.W.; Bieryla, A.; Mann, A.W. Stellar and planetary properties of K2 campaign 1 candidates and validation of 17 planets, including a planet receiving Earth-like insolation. *Astrophys. J.* **2015**, *809*, 25. [CrossRef]
12. Benneke, B.; Wong, I.; Piaulet, C.; Knutson, H.A.; Lothringer, J.; Morley, C.V.; Crossfield, I.J.M.; Gao, P.; Greene, T.P.; Dressing, C. Water vapor and clouds on the habitable-zone sub-Neptune exoplanet K2-18b. *Astrophys. J. Lett.* **2019**, *887*, L14. [CrossRef]
13. Rogers, L.A.; Seager, S. Three possible origins for the gas layer on GJ 1214B. *Astrophys. J.* **2010**, *716*. [CrossRef]
14. Madhusudhan, N.; Nixon, M.C.; Welbanks, L.; Piette, A.A.A.; Booth, R.A. The Interior and Atmosphere of the Habitable-zone Exoplanet K2-18b. *Astrophys. J.* **2020**, *891*, L7. [CrossRef]
15. Greene, T.P.; Line, M.R.; Montero, C.; Fortney, J.J.; Lustig-Yaeger, J.; Luther, K. Characterizing Transiting Exoplanet Atmospheres with JWST. *Astrophys. J.* **2016**, *817*, 17. [CrossRef]
16. Howard, A.W.; Marcy, G.W.; Bryson, S.T.; Jenkins, J.M.; Rowe, J.F.; Batalha, N.M.; Borucki, W.J.; Koch, D.G.; Dunham, E.W.; Gautier, T.N.; et al. Planet occurrence within 0.25AU of solar-type stars from Kepler. *Astrophys. J. Suppl. Ser.* **2012**, *201*. [CrossRef]
17. Dressing, C.D.; Charbonneau, D. The Occurrence Rate of Small Planets around Small Stars. *Astrophys. J.* **2013**, *767*, 95. [CrossRef]
18. Fressin, F.; Torres, G.; Charbonneau, D.; Bryson, S.T.; Christiansen, J.; Dressing, C.D.; Jenkins, J.M.; Walkowicz, L.M.; Batalha, N.M. The False Positive Rate of Kepler and the Occurrence of Planets. *Astrophys. J.* **2013**, *766*, 81. [CrossRef]
19. Bean, J.L.; Raymond, S.N.; Owen, J.E. The Nature and Origins of Sub-Neptune Size Planets. *J. Geophys. Res.* **2021**, *126*, e06639. [CrossRef]
20. Rogers, L.A. Most 1.6 Earth-radius Planets are Not Rocky. *Astrophys. J.* **2015**, *801*, 41. [CrossRef]

21. Kite, E.S.; Fegley, B., Jr.; Schaefer, L.; Ford, E.B. Atmosphere Origins for Exoplanet Sub-Neptunes. *Astrophys. J.* **2020**, *891*, 111. [CrossRef]
22. Vazan, A.; Ormel, C.W.; Noack, L.; Dominik, C. Contribution of the Core to the Thermal Evolution of Sub-Neptunes. *Astrophys. J.* **2018**, *869*, 163. [CrossRef]
23. Bodenheimer, P.; Lissauer, J.J. Accretion and Evolution of ~2.5 M$_\oplus$ Planets with Voluminous H/He Envelopes. *Astrophys. J.* **2014**, *791*, 103. [CrossRef]
24. Bains, W. What do we think life is? A simple illustration and its consequences. *Int. J. Astrobiol.* **2014**, *13*, 101–111. [CrossRef]
25. Baross, J.; Benner, S.A.; Cody, G.D.; Copley, S.D.; Pace, N.R.; Scott, J.H.; Shapiro, R.; Sogin, M.L.; Stein, J.L.; Summons, R.; et al. *The Limits of Organic Life in Planetary Systems*; National Academies Press: Washington, DC, USA, 2007; ISBN 0309179564.
26. Ricker, G.R.; Winn, J.N.; Vanderspek, R.; Latham, D.W.; Bakos, G.Á.; Bean, J.L.; Berta-Thompson, Z.K.; Brown, T.M.; Buchhave, L.; Butler, N.R.; et al. Transiting Exoplanet Survey Satellite (TESS). *J. Astron. Telesc. Instrum. Syst.* **2015**, *1*, 14003. [CrossRef]
27. Vaïtilingom, M.; Attard, E.; Gaiani, N.; Sancelme, M.; Deguillaume, L.; Flossmann, A.I.; Amato, P.; Delort, A.-M. Long-term features of cloud microbiology at the puy de Dôme (France). *Atmos. Environ.* **2012**, *56*, 88–100. [CrossRef]
28. Amato, P.; Joly, M.; Besaury, L.; Oudart, A.; Taib, N.; Moné, A.I.; Deguillaume, L.; Delort, A.-M.; Debroas, D. Active microorganisms thrive among extremely diverse communities in cloud water. *PLoS ONE* **2017**, *12*, e0182869. [CrossRef]
29. Barberán, A.; Ladau, J.; Leff, J.W.; Pollard, K.S.; Menninger, H.L.; Dunn, R.R.; Fierer, N. Continental-scale distributions of dust-associated bacteria and fungi. *Proc. Natl. Acad. Sci. USA* **2015**, *112*, 5756–5761. [CrossRef]
30. Griffin, D.; Gonzalez-Martin, C.; Hoose, C.; Smith, D. Global-scale atmospheric dispersion of microorganisms. *Microbiol. Aerosols* **2017**, 155–194. [CrossRef]
31. Šantl-Temkiv, T.; Gosewinkel, U.; Starnawski, P.; Lever, M.; Finster, K. Aeolian dispersal of bacteria in southwest Greenland: Their sources, abundance, diversity and physiological states. *FEMS Microbiol. Ecol.* **2018**, *94*, fiy031. [CrossRef]
32. Burrows, S.M.; Butler, T.; Jöckel, P.; Tost, H.; Kerkweg, A.; Pöschl, U.; Lawrence, M.G. Bacteria in the global atmosphere–Part 2: Modeling of emissions and transport between different ecosystems. *Atmos. Chem. Phys.* **2009**, *9*, 9281–9297. [CrossRef]
33. Bowers, R.M.; McCubbin, I.B.; Hallar, A.G.; Fierer, N. Seasonal variability in airborne bacterial communities at a high-elevation site. *Atmos. Environ.* **2012**, *50*, 41–49. [CrossRef]
34. Amato, P.; Ménager, M.; Sancelme, M.; Laj, P.; Mailhot, G.; Delort, A.-M. Microbial population in cloud water at the Puy de Dôme: Implications for the chemistry of clouds. *Atmos. Environ.* **2005**, *39*, 4143–4153. [CrossRef]
35. DeLong, E.F. Diminutive Cells in The Oceans—Unanswered Questions. In *Size Limits of Very Small Microorganisms: Proceedings of a Workshop*; Monterey Bay Aquarium Research Institute: Moss Landing, CA, USA, 1999; p. 81.
36. Seager, S.; Petkowski, J.J.; Gao, P.; Bains, W.; Bryan, N.C.; Ranjan, S.; Greaves, J. The Venusian Lower Atmosphere Haze as a Depot for Desiccated Microbial Life: A Proposed Life Cycle for Persistence of the Venusian Aerial Biosphere. *Astrobiology* **2020**. [CrossRef]
37. Bryan, N.C.; Christner, B.C.; Guzik, T.G.; Granger, D.J.; Stewart, M.F. Abundance and survival of microbial aerosols in the troposphere and stratosphere. *ISME J.* **2019**, *13*, 1–11. [CrossRef] [PubMed]
38. Amato, P.; Besaury, L.; Joly, M.; Penaud, B.; Deguillaume, L.; Delort, A.-M. Metatranscriptomic exploration of microbial functioning in clouds. *Sci. Rep.* **2019**, *9*, 4383. [CrossRef]
39. Dimmick, R.L.; Wolochow, H.; Chatigny, M.A. Evidence for more than one division of bacteria within airborne particles. *Appl. Environ. Microbiol.* **1979**, *38*, 642–643. [CrossRef]
40. Dimmick, R.L.; Wolochow, H.; Chatigny, M.A. Evidence that bacteria can form new cells in airborne particles. *Appl. Environ. Microbiol.* **1979**, *37*, 924–927. [CrossRef]
41. Sattler, B.; Puxbaum, H.; Psenner, R. Bacterial growth in supercooled cloud droplets. *Geophys. Res. Lett.* **2001**, *28*, 239–242. [CrossRef]
42. Morowitz, H.; Sagan, C. Life in the Clouds of Venus? *Nature* **1967**, *215*, 1259. [CrossRef]
43. Limaye, S.S.; Mogul, R.; Smith, D.J.; Ansari, A.H.; Słowik, G.P.; Vaishampayan, P. Venus' Spectral Signatures and the Potential for Life in the Clouds. *Astrobiology* **2018**, *18*, 1181–1198. [CrossRef]
44. Cavalazzi, B.; Barbieri, R.; Gómez, F.; Capaccioni, B.; Olsson-Francis, K.; Pondrelli, M.; Rossi, A.P.; Hickman-Lewis, K.; Agangi, A.; Gasparotto, G.; et al. The Dallol Geothermal Area, Northern Afar (Ethiopia)-An Exceptional Planetary Field Analog on Earth. *Astrobiology* **2019**, *19*, 553–578. [CrossRef]
45. Kotopoulou, E.; Delgado Huertas, A.; Garcia-Ruiz, J.M.; Dominguez-Vera, J.M.; Lopez-Garcia, J.M.; Guerra-Tschuschke, I.; Rull, F. A polyextreme hydrothermal system controlled by iron: The case of Dallol at the Afar Triangle. *ACS Earth Sp. Chem.* **2018**, *3*, 90–99. [CrossRef] [PubMed]
46. Sagan, C.; Salpeter, E.E. Particles, environments and possible ecologies in the Jovian atmosphere. *Astrophys. J. Suppl. Ser.* **1976**, *32*, 737–755. [CrossRef]
47. Yates, J.S.; Palmer, P.I.; Biller, B.; Cockell, C.S. Atmospheric habitable zones in Y dwarf atmospheres. *Astrophys. J.* **2017**, *836*, 184. [CrossRef]
48. Lingam, M.; Loeb, A. Brown Dwarf atmospheres as the potentially most detectable and abundant sites for life. *Astrophys. J.* **2019**, *883*, 143. [CrossRef]
49. Tsiaras, A.; Waldmann, I.P.; Tinetti, G.; Tennyson, J.; Yurchenko, S.N. Water vapour in the atmosphere of the habitable-zone eight-Earth-mass planet K2-18 b. *Nat. Astron.* **2019**, *3*, 1086–1091. [CrossRef]

50. Blain, D.; Charnay, B.; Bézard, B. 1D atmospheric study of the temperate sub-Neptune K2-18b. *Astron. Astrophys.* **2021**, *646*, A15. [CrossRef]
51. Charnay, B.; Blain, D.; Bézard, B.; Leconte, J.; Turbet, M.; Falco, A. Formation and dynamics of water clouds on temperate sub-Neptunes: The example of K2-18b. *Astron. Astrophys.* **2021**, *646*, A171. [CrossRef]
52. Bains, W.; Xiao, Y.; Yu, C. Prediction of the maximum temperature for life based on the stability of metabolites to decomposition in water. *Life* **2015**, *5*, 1054–1100. [CrossRef]
53. McKay, C.P. Requirements and limits for life in the context of exoplanets. *Proc. Natl. Acad. Sci. USA* **2014**, *111*, 12628–12633. [CrossRef]
54. Takai, K.; Nakamura, K.; Toki, T.; Tsunogai, U.; Miyazaki, M.; Miyazaki, J.; Hirayama, H.; Nakagawa, S.; Nunoura, T.; Horikoshi, K. Cell proliferation at 122 C and isotopically heavy $CH_4$ production by a hyperthermophilic methanogen under high-pressure cultivation. *Proc. Natl. Acad. Sci. USA* **2008**, *105*, 10949–10954. [CrossRef]
55. Kashefi, K.; Lovley, D.R. Extending the upper temperature limit for life. *Science* **2003**, *301*, 934. [CrossRef]
56. Hoehler, T.; Bains, W.; Davila, A.; Parenteau, M.; Pohorille, A. Life's requirements, habitability, and biological potential. In *Planetary Astrobiology*; Meadows, V., Marais, D.J., Des Arney, G., Schmidt, B., Eds.; The University of Arizona Press: Tucson, AZ, USA, 2020.
57. Petkowski, J.J.; Bains, W.; Seager, S. On the Potential of Silicon as a Building Block for Life. *Life* **2020**, *10*, 84. [CrossRef] [PubMed]
58. Seager, S.; Huang, J.; Petkowski, J.J.; Pajusalu, M. Laboratory studies on the viability of life in $H_2$-dominated exoplanet atmospheres. *Nat. Astron.* **2020**. [CrossRef]
59. Engel, T.; Reid, P. *Thermodynamics*, 4th ed.; Pearson: New York, NY, USA, 2018; ISBN 978-0134804583.
60. Lodders, K.; Fegley, B. *The Planetary Scientist's Companion/Katharina Lodders, Bruce Fegley*; Oxford University Press: Oxford, UK, 1998.
61. Piette, A.A.A.; Madhusudhan, N. On the Temperature Profiles and Emission Spectra of Mini-Neptune Atmospheres. *Astrophys. J.* **2020**, *904*, 154. [CrossRef]
62. Pierrehumbert, R.T.; Hammond, M. Atmospheric Circulation of Tide-Locked Exoplanets. *Annu. Rev. Fluid Mech.* **2019**, *51*, 275–303. [CrossRef]
63. Guillot, T. Condensation of methane, ammonia, and water and the inhibition of convection in giant planets. *Science* **1995**, *269*, 1697–1699. [CrossRef] [PubMed]
64. Leconte, J.; Selsis, F.; Hersant, F.; Guillot, T. Condensation-inhibited convection in hydrogen-rich atmospheres. *Astron. Astrophys.* **2017**, *598*, A98. [CrossRef]
65. Fievet, A.; Ducret, A.; Mignot, T.; Valette, O.; Robert, L.; Pardoux, R.; Dolla, A.R.; Aubert, C. Single-cell analysis of growth and cell division of the anaerobe Desulfovibrio vulgaris Hildenborough. *Front. Microbiol.* **2015**, *6*, 1378. [CrossRef] [PubMed]
66. Goldreich, P.; Soter, S. Q in the Solar System. *Icarus* **1966**, *5*, 375–389. [CrossRef]
67. Guinan, E.F.; Engle, S.G. The K2-18b Planetary System: Estimates of the Age and X-UV Irradiances of a Habitable Zone "Wet" Sub-Neptune Planet. *Res. Notes Am. Astron. Soc.* **2019**, *3*, 189. [CrossRef]
68. Kong, D.; Zhang, K.; Schubert, G.; Anderson, J.D. Origin of Jupiter's cloud-level zonal winds remains a puzzle even after Juno. *Proc. Natl. Acad. Sci. USA* **2018**, *115*, 8499–8504. [CrossRef]
69. Fletcher, L.N.; Kaspi, Y.; Guillot, T.; Showman, A.P. How Well Do We Understand the Belt/Zone Circulation of Giant Planet Atmospheres? *Space Sci. Rev.* **2020**, *216*, 30. [CrossRef]
70. Showman, A.P.; Tan, X.; Parmentier, V. Atmospheric Dynamics of Hot Giant Planets and Brown Dwarfs. *Space Sci. Rev.* **2020**, *216*, 139. [CrossRef]
71. West, R.A.; Friedson, A.J.; Appleby, J.F. Jovian large-scale stratospheric circulation. *Icarus* **1992**, *100*, 245–259. [CrossRef]
72. Walters, K.R.; Serianni, A.S.; Sformo, T.; Barnes, B.M.; Duman, J.G. A nonprotein thermal hysteresis-producing xylomannan antifreeze in the freeze-tolerant Alaskan beetle Upis ceramboides. *Proc. Natl. Acad. Sci. USA* **2009**, *106*, 20210–20215. [CrossRef] [PubMed]
73. Mykytczuk, N.C.S.; Foote, S.J.; Omelon, C.R.; Southam, G.; Greer, C.W.; Whyte, L.G. Bacterial growth at 15 °C; molecular insights from the permafrost bacterium Planococcus halocryophilus Or1. *ISME J.* **2013**, *7*, 1211–1226. [CrossRef]
74. Maykut, G.; Untersteiner, N. *The Geophysics of Sea Ice*; Plenium: New York, NY, USA, 1986.
75. Junge, K.; Eicken, H.; Deming, J.W. Motility of Colwellia psychrerythraea strain 34H at subzero temperatures. *Appl. Environ. Microbiol.* **2003**, *69*, 4282–4284. [CrossRef]
76. Huston, A.L. Bacterial Adaptation to the Cold: In Situ Activities of Extracellular Enzymes in the North Water Polynya and Characterization of a Cold-Active Aminopeptidase from Colwellia Psychrerythraea Strain 34H. Ph.D. Thesis, University of Washington, Washington, DC, USA, 2004.
77. Farrell, A.H.; Hohenstein, K.A.; Shain, D.H. Molecular adaptation in the ice worm, Mesenchytraeus solifugus: Divergence of energetic-associated genes. *J. Mol. Evol.* **2004**, *59*, 666–673. [CrossRef]
78. Shain, D.H.; Mason, T.A.; Farrell, A.H.; Michalewicz, L.A. Distribution and behavior of ice worms (Mesenchytraeus solifugus) in south-central Alaska. *Can. J. Zool.* **2001**, *79*, 1813–1821. [CrossRef]
79. Shain, D.H.; Carter, M.R.; Murray, K.P.; Maleski, K.A.; Smith, N.R.; McBride, T.R.; Michalewicz, L.A.; Saidel, W.M. Morphologic characterization of the ice worm Mesenchytraeus solifugus. *J. Morphol.* **2000**, *246*, 192–197. [CrossRef]

80. Cáceres, L.; Gómez-Silva, B.; Garró, X.; Rodríguez, V.; Monardes, V.; McKay, C.P. Relative humidity patterns and fog water precipitation in the Atacama Desert and biological implications. *J. Geophys. Res. Biogeosciences* **2007**, *112*. [CrossRef]
81. Gruzdev, N.; McClelland, M.; Porwollik, S.; Ofaim, S.; Pinto, R.; Saldinger-Sela, S. Global Transcriptional Analysis of Dehydrated Salmonella enterica Serovar Typhimurium. *Appl. Environ. Microbiol.* **2012**, *78*, 7866–7875. [CrossRef]
82. Alpert, P. The limits and frontiers of desiccation-tolerant life. *Integr. Comp. Biol.* **2005**, *45*, 685–695. [CrossRef]
83. Guidetti, R.; JoÈnsson, K.I. Long-term anhydrobiotic survival in semi-terrestrial micrometazoans. *J. Zool.* **2002**, *257*, 181–187. [CrossRef]
84. Nicholson, W.L.; Munakata, N.; Horneck, G.; Melosh, H.J.; Setlow, P. Resistance of Bacillus endospores to extreme terrestrial and extraterrestrial environments. *Microbiol. Mol. Biol. Rev.* **2000**, *64*, 548–572. [CrossRef] [PubMed]
85. Christner, B.C.; Mosley-Thompson, E.; Thompson, L.G.; Zagorodnov, V.; Sandman, K.; Reeve, J.N. Recovery and Identification of Viable Bacteria Immured in Glacial Ice. *Icarus* **2000**, *144*, 479–485. [CrossRef]
86. Cano, R.J.; Borucki, M.K. Revival and identification of bacterial spores in 25- to 40-million-year-old Dominican amber. *Science* **1995**, *268*, 1060–1064. [CrossRef]
87. Meng, F.-W.; Wang, X.-Q.; Ni, P.; Kletetschka, G.; Yang, C.-H.; Li, Y.-P.; Yang, Y.-H. A newly isolated Haloalkaliphilic bacterium from middle–late Eocene halite formed in salt lakes in China. *Carbonates Evaporites* **2015**, *30*, 321–330. [CrossRef]
88. Burch, A.Y.; Zeisler, V.; Yokota, K.; Schreiber, L.; Lindow, S.E. The hygroscopic biosurfactant syringafactin produced by Pseudomonas syringae enhances fitness on leaf surfaces during fluctuating humidity. *Environ. Microbiol.* **2014**, *16*, 2086–2098. [CrossRef]
89. Elliot, M.A.; Talbot, N.J. Building filaments in the air: Aerial morphogenesis in bacteria and fungi. *Curr. Opin. Microbiol.* **2004**, *7*, 594–601. [CrossRef]
90. Talbot, N.J. Coming up for air and sporulation. *Nature* **1999**, *398*, 295–296. [CrossRef]
91. Hedenström, A.; Norevik, G.; Warfvinge, K.; Andersson, A.; Bäckman, J.; Åkesson, S. Annual 10-month aerial life phase in the common swift Apus apus. *Curr. Biol.* **2016**, *26*, 3066–3070. [CrossRef]
92. Lyons, T.W.; Rogers, K.; Krishnamurthy, R.; Williams, L.; Marchi, S.; Schwieterman, E.; Planavsky, N.; Reinhard, C. Constraining Prebiotic Chemistry Through a Better Understanding of Earth's Earliest Environments. *arXiv* **2020**, arXiv:2008.04803.
93. Bains, W. Getting Beyond the Toy Domain. Meditations on David Deamer's "Assembling Life". *Life* **2020**, *10*, 18. [CrossRef] [PubMed]
94. Shapiro, R. *Origins: A Skeptic's Guide to the Creation of Life on Earth*; Bantam Dell Pub Group: New York, NY, USA, 1987; ISBN 0553343556.
95. Benner, S.A.; Bell, E.A.; Biondi, E.; Brasser, R.; Carell, T.; Kim, H.; Mojzsis, S.J.; Omran, A.; Pasek, M.A.; Trail, D. When Did Life Likely Emerge on Earth in an RNA-First Process? *ChemSystemsChem* **2019**, *2*, e1900035.
96. Furukawa, Y.; Chikaraishi, Y.; Ohkouchi, N.; Ogawa, N.O.; Glavin, D.P.; Dworkin, J.P.; Abe, C.; Nakamura, T. Extraterrestrial ribose and other sugars in primitive meteorites. *Proc. Natl. Acad. Sci. USA* **2019**, *116*, 24440–24445. [CrossRef]
97. Callahan, M.P.; Smith, K.E.; Cleaves, H.J.; Ruzicka, J.; Stern, J.C.; Glavin, D.P.; House, C.H.; Dworkin, J.P. Carbonaceous meteorites contain a wide range of extraterrestrial nucleobases. *Proc. Natl. Acad. Sci. USA* **2011**, *108*, 13995–13998. [CrossRef] [PubMed]
98. Dong, J.; Fischer, R.A.; Stixrude, L.P.; Lithgow-Bertelloni, C.R. Constraining the Volume of Earth's Early Oceans With a Temperature-Dependent Mantle Water Storage Capacity Model. *AGU Adv.* **2021**, *2*, e2020AV000323. [CrossRef]
99. Woese, C.R. A proposal concerning the origin of life on the planet earth. *J. Mol. Evol.* **1979**, *13*, 95–101. [CrossRef] [PubMed]
100. Damer, B.; Deamer, D. Coupled phases and combinatorial selection in fluctuating hydrothermal pools: A scenario to guide experimental approaches to the origin of cellular life. *Life* **2015**, *5*, 872–887. [CrossRef] [PubMed]
101. Becker, S.; Schneider, C.; Okamura, H.; Crisp, A.; Amatov, T.; Dejmek, M.; Carell, T. Wet-dry cycles enable the parallel origin of canonical and non-canonical nucleosides by continuous synthesis. *Nat. Commun.* **2018**, *9*, 1–9. [CrossRef]
102. Banerjee, S.; Gnanamani, E.; Yan, X.; Zare, R.N. Can all bulk-phase reactions be accelerated in microdroplets? *Analyst* **2017**, *142*, 1399–1402. [CrossRef]
103. Nam, I.; Lee, J.K.; Nam, H.G.; Zare, R.N. Abiotic production of sugar phosphates and uridine ribonucleoside in aqueous microdroplets. *Proc. Natl. Acad. Sci. USA* **2017**, *114*, 12396–12400. [CrossRef] [PubMed]
104. Griffith, E.C.; Vaida, V. In situ observation of peptide bond formation at the water-air interface. *Proc. Natl. Acad. Sci. USA* **2012**, *109*, 15697–15701. [CrossRef]
105. Toupance, G.; Raulin, F.; Buvet, R. Formation of prebiochemical compounds in models of the primitive Earth's atmosphere. *Orig. Life* **1975**, *6*, 83–90. [CrossRef]
106. Shi, W.; Chance, M.R. Metalloproteomics: Forward and reverse approaches in metalloprotein structural and functional characterization. *Curr. Opin. Chem. Biol.* **2011**, *15*, 144–148. [CrossRef]
107. Peucker-Ehrenbrink, B. Accretion of extraterrestrial matter during the last 80 million years and its effect on the marine osmium isotope record. *Geochim. Cosmochim. Acta* **1996**, *60*, 3187–3196. [CrossRef]
108. Emiliani, C. *Planet. Earth: Cosmology, Geology, and the Evolution of Life and Environment*; Cambridge University Press: Cambridge, UK, 1992; ISBN 9780521401234.
109. Fung, I.Y.; Meyn, S.K.; Tegen, I.; Doney, S.C.; John, J.G.; Bishop, J.K.B. Iron supply and demand in the upper ocean. *Glob. Biogeochem. Cycles* **2000**, *14*, 281–295. [CrossRef]

110. Li, J.; Mara, P.; Schubotz, F.; Sylvan, J.B.; Burgaud, G.; Klein, F.; Beaudoin, D.; Wee, S.Y.; Dick, H.J.B.; Lott, S. Recycling and metabolic flexibility dictate life in the lower oceanic crust. *Nature* **2020**, *579*, 250–255. [CrossRef]
111. Oza, A.V.; Johnson, R.E.; Lellouch, E.; Schmidt, C.; Schneider, N.; Huang, C.; Gamborino, D.; Gebek, A.; Wyttenbach, A.; Demory, B.-O.; et al. Sodium and Potassium Signatures of Volcanic Satellites Orbiting Close-in Gas Giant Exoplanets. *Astrophys. J.* **2019**, *885*, 168. [CrossRef]
112. Thomas, N.; Bagenal, F.; Hill, T.W.; Wilson, J.K. The Io neutral clouds and plasma torus. *Jupit. Planet Satell. Magnetos.* **2004**, *1*, 561–591.
113. Tucker, O.J.; Johnson, R.E.; Young, L.A. Gas transfer in the Pluto-Charon system: A Charon atmosphere. *Icarus* **2015**, *246*, 291–297. [CrossRef]
114. Heller, R.; Williams, D.; Kipping, D.; Limbach, M.A.; Turner, E.; Greenberg, R.; Sasaki, T.; Bolmont, É.; Grasset, O.; Lewis, K.; et al. Formation, Habitability, and Detection of Extrasolar Moons. *Astrobiology* **2014**, *14*, 798–835. [CrossRef]
115. Hoehler, T.M. Biological energy requirements as quantitative boundary conditions for life in the subsurface. *Geobiology* **2004**, *2*, 205–215. [CrossRef]
116. Ranjan, S.; Wordsworth, R.; Sasselov, D.D. The surface UV environment on planets orbiting M dwarfs: Implications for prebiotic chemistry and the need for experimental follow-up. *Astrophys. J.* **2017**, *843*, 110. [CrossRef]
117. Rimmer, P.B.; Xu, J.; Thompson, S.J.; Gillen, E.; Sutherland, J.D.; Queloz, D. The origin of RNA precursors on exoplanets. *Sci. Adv.* **2018**, *4*, eaar3302. [CrossRef] [PubMed]
118. Günther, M.N.; Zhan, Z.; Seager, S.; Rimmer, P.B.; Ranjan, S.; Stassun, K.G.; Oelkers, R.J.; Daylan, T.; Newton, E.; Kristiansen, M.H.; et al. Stellar Flares from the First TESS Data Release: Exploring a New Sample of M Dwarfs. *Astron. J.* **2020**, *159*, 60. [CrossRef]
119. Segura, A.; Walkowicz, L.M.; Meadows, V.; Kasting, J.; Hawley, S. The effect of a strong stellar flare on the atmospheric chemistry of an Earth-like planet orbiting an M dwarf. *Astrobiology* **2010**, *10*, 751–771. [CrossRef] [PubMed]
120. MacGregor, M.A.; Weinberger, A.J.; Loyd, R.O.P.; Shkolnik, E.; Barclay, T.; Howard, W.S.; Zic, A.; Osten, R.A.; Cranmer, S.R.; Kowalski, A.F.; et al. Discovery of an Extremely Short Duration Flare from Proxima Centauri Using Millimeter through Far-ultraviolet Observations. *Astrophys. J.* **2021**, *911*, L25. [CrossRef]
121. Ranjan, S.; Sasselov, D.D. Influence of the UV Environment on the Synthesis of Prebiotic Molecules. *Astrobiology* **2016**, *16*, 68–88. [CrossRef] [PubMed]
122. Cox, M.M.; Keck, J.L.; Battista, J.R. Rising from the Ashes: DNA Repair in Deinococcus radiodurans. *PLoS Genet.* **2010**, *6*, e1000815. [CrossRef] [PubMed]
123. Diaz, B.; Schulze-Makuch, D. Microbial Survival Rates of Escherichia coli and Deinococcus radiodurans Under Low Temperature, Low Pressure, and UV-Irradiation Conditions, and Their Relevance to Possible Martian Life. *Astrobiology* **2006**, *6*, 332–347. [CrossRef] [PubMed]
124. Hawley, S.L.; Davenport, J.R.A.; Kowalski, A.F.; Wisniewski, J.P.; Hebb, L.; Deitrick, R.; Hilton, E.J. Kepler Flares. I. Active and Inactive M Dwarfs. *Astrophys. J.* **2014**, *797*, 121. [CrossRef]
125. Newton, E.R.; Irwin, J.; Charbonneau, D.; Berlind, P.; Calkins, M.L.; Mink, J. The H$\alpha$ Emission of Nearby M Dwarfs and its Relation to Stellar Rotation. *Astrophys. J.* **2017**, *834*, 85. [CrossRef]
126. Owen, J.E.; Wu, Y. Kepler Planets: A Tale of Evaporation. *Astrophys. J.* **2013**, *775*, 105. [CrossRef]
127. Mordasini, C. Planetary evolution with atmospheric photoevaporation. I. Analytical derivation and numerical study of the evaporation valley and transition from super-Earths to sub-Neptunes. *Astron. Astrophys.* **2020**, *638*, A52. [CrossRef]
128. Atri, D.; Mogan, S.R.C. Stellar flares versus luminosity: XUV-induced atmospheric escape and planetary habitability. *Mon. Not. R. Astron. Soc.* **2021**, *500*, L1–L5. [CrossRef]
129. Seager, S.; Bains, W.; Petkowski, J.J. Toward a List of Molecules as Potential Biosignature Gases for the Search for Life on Exoplanets and Applications to Terrestrial Biochemistry. *Astrobiology* **2016**, *16*, 465–485. [CrossRef] [PubMed]
130. Schwieterman, E.W.; Kiang, N.Y.; Parenteau, M.N.; Harman, C.E.; DasSarma, S.; Fisher, T.M.; Arney, G.N.; Hartnett, H.E.; Reinhard, C.T.; Olson, S.L.; et al. Exoplanet Biosignatures: A Review of Remotely Detectable Signs of Life. *Astrobiology* **2018**, *18*, 663–708. [CrossRef]
131. Seager, S.; Schrenk, M.; Bains, W. An Astrophysical View of Earth-Based Metabolic Biosignature Gases. *Astrobiology* **2012**, *12*, 61–82. [CrossRef] [PubMed]
132. Haas, J.R. The potential feasibility of chlorinic photosynthesis on exoplanets. *Astrobiology* **2010**, *10*, 953–963. [CrossRef]
133. Seager, S.; Bains, W.; Hu, R. Biosignature Gases in $H_2$-dominated Atmospheres on Rocky Exoplanets. *Astrophys. J.* **2013**, *777*, 95. [CrossRef]
134. Bains, W.; Seager, S.; Zsom, A. Photosynthesis in hydrogen-dominated atmospheres. *Life* **2014**, *4*, 716–744. [CrossRef]
135. Valmalette, J.C.; Dombrovsky, A.; Brat, P.; Mertz, C.; Capovilla, M.; Robichon, A. Light-induced electron transfer and ATP synthesis in a carotene synthesizing insect. *Sci. Rep.* **2012**, *2*, 579. [CrossRef]
136. Moses, J.I.; Line, M.R.; Visscher, C.; Richardson, M.R.; Nettelmann, N.; Fortney, J.J.; Barman, T.S.; Stevenson, K.B.; Madhusudhan, N. Compositional Diversity in the Atmospheres of Hot Neptunes, with Application to GJ 436B. *Astrophys. J.* **2013**, *777*, 34. [CrossRef] [PubMed]
137. Hoehler, T.M.; Alperin, M.J.; Albert, D.B.; Martens, C.S. Apparent minimum free energy requirements for methanogenic Archaea and sulfate-reducing bacteria in an anoxic marine sediment. *FEMS Microbiol. Ecol.* **2001**, *38*, 33–41. [CrossRef]

138. Hu, R.; Seager, S.; Bains, W. Photochemistry in terrestrial exoplanet atmospheres. I. Photochemistry model and benchmark cases. *Astrophys. J.* **2012**, *761*, 166. [CrossRef]
139. Mollière, P.; Snellen, I.A.G. Detecting isotopologues in exoplanet atmospheres using ground-based high-dispersion spectroscopy. *Astron. Astrophys.* **2019**, *622*, A139. [CrossRef]
140. Arnold, L.; Gillet, S.; Lardière, O.; Riaud, P.; Schneider, J. A test for the search for life on extrasolar planets. *Astron. Astrophys.* **2002**, *392*, 231–237. [CrossRef]
141. Woolf, N.J.; Smith, P.S.; Traub, W.A.; Jucks, K.W. The Spectrum of Earthshine: A Pale Blue Dot Observed from the Ground. *Astrophys. J.* **2002**, *574*, 430–433. [CrossRef]
142. Seager, S.; Turner, E.L.; Schafer, J.; Ford, E.B. Vegetation's red edge: A possible spectroscopic biosignature of extraterrestrial plants. *Astrobiology* **2005**, *5*. [CrossRef] [PubMed]
143. Montanes-Rodriguez, P.; Palle, E.; Goode, P.R.; Martin-Torres, F.J. Vegetation Signature in the Observed Globally Integrated Spectrum of Earth Considering Simultaneous Cloud Data: Applications for Extrasolar Planets. *Astrophys. J.* **2006**, *651*, 544–552. [CrossRef]
144. Turnbull, M.C.; Traub, W.A.; Jucks, K.W.; Woolf, N.J.; Meyer, M.R.; Gorlova, N.; Skrutskie, M.F.; Wilson, J.C. Spectrum of a Habitable World: Earthshine in the Near-Infrared. *Astrophys. J.* **2006**, *644*, 551–559. [CrossRef]
145. Kiang, N.Y.; Segura, A.; Tinetti, G.; Govindjee; Blankenship, R.E.; Cohen, M.; Siefert, J.; Crisp, D.; Meadows, V.S. Spectral Signatures of Photosynthesis. II. Coevolution with Other Stars And The Atmosphere on Extrasolar Worlds. *Astrobiology* **2007**, *7*, 252–274. [CrossRef]
146. Hegde, S.; Paulino-Lima, I.G.; Kent, R.; Kaltenegger, L.; Rothschild, L. Surface biosignatures of exo-Earths: Remote detection of extraterrestrial life. *Proc. Natl. Acad. Sci. USA* **2015**, *112*, 3886–3891. [CrossRef]
147. Kempton, E.M.; Bean, J.L.; Louie, D.R.; Deming, D.; Koll, D.D.B.; Mansfield, M.; Christiansen, J.L.; López-Morales, M.; Swain, M.R.; Zellem, R.T.; et al. A framework for prioritizing the TESS planetary candidates most amenable to atmospheric characterization. *Publ. Astron. Soc. Pac.* **2018**, *130*. [CrossRef]
148. Seager, S. *Exoplanet Atmospheres: Physical Processes*; Princeton University Press: Princeton, NJ, USA, 2010; ISBN 9780691130262.
149. Weiss, L.M.; Marcy, G.W. The Mass-Radius Relation for 65 Exoplanets Smaller than 4 Earth Radii. *Astrophys. J.* **2014**, *783*, L6. [CrossRef]
150. Chen, J.; Kipping, D. Probabilistic Forecasting of the Masses and Radii of Other Worlds. *Astrophys. J.* **2017**, *834*, 17. [CrossRef]